\ifx\documentstyle\undefined\else \fi 
\xdef\fmtversion{\fmtversion+CWEB3.1}

\let\:=\. 

\parskip 0pt 
\parindent 1em 

\let\mc=\ninerm 
\def\CEE/{{\mc C\spacefactor1000}}
\def\UNIX/{{\mc U\kern-.05emNIX\spacefactor1000}}
\def\TEX/{\TeX}
\def\CPLUSPLUS/{{\mc C\PP\spacefactor1000}}
\def\9#1{}
\font\eightrm=cmr8
\let\sc=\eightrm 
\let\mainfont=\tenrm
\let\cmntfont\tenrm
\font\titlefont=cmr7 scaled\magstep4 
\font\tentex=cmtex10 
\fontdimen7\tentex=0pt 

\def\\#1{\leavevmode\hbox{\it#1\/\kern.05em}} 
\def\|#1{\leavevmode\hbox{$#1$}} 
\def\&#1{\leavevmode\hbox{\bf
  \def\_{\kern.04em\vbox{\hrule width.3em height .6pt}\kern.08em}%
  #1\/\kern.05em}} 
\def\.#1{\leavevmode\hbox{\tentex 
  \let\\=\BS 
  \let\{=\LB 
  \let\}=\RB 
  \let\~=\TL 
  \let\ =\SP 
  \let\_=\UL 
  \let\&=\AM 
  \let\^=\CF 
  #1\kern.05em}}
\def\){\discretionary{\hbox{\tentex\BS}}{}{}}
\def\ATL{\par\noindent\bgroup\catcode`\_=12 \postATL} 
\def\postATL#1 #2 {\bf letter \\{\uppercase{\char"#1}}
   tangles as \tentex "#2"\egroup\par}
\def\noATL#1 #2 {}
\def\noatl{\let\ATL=\noATL} 
\def\ATH{\X\kern-.5em:Preprocessor definitions\X}
\let\PB=\relax 

\chardef\AM=`\& 
\chardef\BS=`\\ 
\chardef\LB=`\{ 
\chardef\RB=`\} 
\def\SP{{\tt\char`\ }} 
\chardef\TL=`\~ 
\chardef\UL=`\_ 
\chardef\CF=`\^ 

\newbox\PPbox 
\setbox\PPbox=\hbox{\kern.5pt\raise1pt\hbox{\sevenrm+\kern-1pt+}\kern.5pt}
\def\PP{\copy\PPbox}
\newbox\MMbox \setbox\MMbox=\hbox{\kern.5pt\raise1pt\hbox{\sevensy\char0
 \kern-1pt\char0}\kern.5pt}

\newbox\MGbox 
\setbox\MGbox=\hbox{\kern-2pt\lower3pt\hbox{\teni\char'176}\kern1pt}
\def\MG{\copy\MGbox}
\def\MRL#1{\mathrel{\let\K==#1}}

\let\NULL=\Lambda
\mathchardef\AND="2026 
\newbox\MODbox \setbox\MODbox=\hbox{\eightrm\%}
\def\MOD{\mathbin{\copy\MODbox}}
\def\DC{\kern.1em{::}\kern.1em} 
\def\this{\&{this}}

\newbox\bak \setbox\bak=\hbox to -1em{} 
\newbox\bakk\setbox\bakk=\hbox to -2em{} 

\newcount\ind 
\def\1{\global\advance\ind by1\hangindent\ind em} 
\def\2{\global\advance\ind by-1} 
\def\3#1{\hfil\penalty#10\hfilneg} 
\def\4{\copy\bak} 
\def\5{\hfil\penalty-1\hfilneg\kern2.5em\copy\bakk\ignorespaces}
\def\6{\ifmmode\else\par 
  \hangindent\ind em\noindent\kern\ind em\copy\bakk\ignorespaces\fi}
\def\7{\Y\6} 
\def\8{\hskip-\ind em\hskip 2em} 

\newcount\gdepth 
\newcount\secpagedepth
\secpagedepth=3 
\newtoks\gtitle 
\newskip\intersecskip \intersecskip=12pt minus 3pt 
\let\yskip=\smallskip
\def\?{\mathrel?}
\def\note#1#2.{\Y\noindent{\hangindent2em\baselineskip10pt\eightrm#1~#2.\par}}
\def\lapstar{\rlap{*}}
\def\stsec{\rightskip=0pt 
  \sfcode`;=1500 \pretolerance 200 \hyphenpenalty 50 \exhyphenpenalty 50
  \noindent{\let\*=\lapstar\bf\secstar.\quad}}
\let\startsection=\stsec
\def\defin#1{\global\advance\ind by 2 \1\&{#1 } } 
\def\B{\rightskip=0pt plus 100pt minus 10pt 
  \sfcode`;=3000
  \pretolerance 10000
  \hyphenpenalty 1000 
  \exhyphenpenalty 10000
  \global\ind=2 \1\ \unskip}
\def\C#1{\5\5\quad$/\ast\,${\cmntfont #1}$\,\ast/$}
\def\D{\defin{\#define}} 
\let\E=\equiv 
\def\ET{ and~} 
\def\ETs{, and~} 
\let\G=\ge 
\let\I=\ne 
\let\K== 
\outer\def\M#1{\MN{#1}\ifon\vfil\penalty-100\vfilneg 
  \vskip\intersecskip\startsection\ignorespaces}
\outer\def\N#1#2#3.{\gdepth=#1\gtitle={#3}\MN{#2}
  \ifon\ifnum#1<\secpagedepth \vfil\eject 
    \else\vfil\penalty-100\vfilneg\vskip\intersecskip\fi\fi
  \message{*\secno} 
  \edef\next{\write\cont{\ZZ{#3}{#1}{\secno}
                   {\noexpand\the\pageno}}}\next 
  \ifon\startsection{\bf#3.\quad}\ignorespaces}
\def\MN#1{\par 
  {\xdef\secstar{#1}\let\*=\empty\xdef\secno{#1}}
  \ifx\secno\secstar \onmaybe \else\ontrue \fi
  \mark{{{\tensy x}\secno}{\the\gdepth}{\the\gtitle}}}
\def\Q{\note{This code is cited in section}} 
\def\Qs{\note{This code is cited in sections}} 
\let\R=\lnot 
\def\T#1{\leavevmode 
  \hbox{$\def\?{\kern.2em}%
    \def\$##1{\egroup_{\,\rm##1}\bgroup}
    \def\_{\cdot 10^{\aftergroup}}
    \let\~=\oct \let\^=\hex {#1}$}}
\def\U{\note{This code is used in section}} 
\def\Us{\note{This code is used in sections}} 
\let\W=\land 
\def\X#1:#2\X{\ifmmode\gdef\XX{\null$\null}\else\gdef\XX{}\fi 
  \XX$\langle\,${#2\eightrm\kern.5em#1}$\,\rangle$\XX}
\def\Y{\par\yskip}
\let\Z=\le
\let\ZZ=\let 
\let\*=*

\def\oct{\hbox{$^\circ$\kern-.1em\it\aftergroup\?\aftergroup}}
\def\hex{\hbox{$^{\scriptscriptstyle\#}$\tt\aftergroup}} 
\def\vb#1{\leavevmode\hbox{\kern2pt\vrule\vtop{\vbox{\hrule
        \hbox{\strut\kern2pt\.{#1}\kern2pt}}
      \hrule}\vrule\kern2pt}} 

\def\onmaybe{\let\ifon=\maybe} \let\maybe=\iftrue
\newif\ifon \newif\iftitle \newif\ifpagesaved

\def\lheader{\mainfont\the\pageno\eightrm\qquad\grouptitle\hfill\title\qquad
  \mainfont\topsecno} 
\def\rheader{\mainfont\topsecno\eightrm\qquad\title\hfill\grouptitle
  \qquad\mainfont\the\pageno} 
\def\grouptitle{\let\i=I\let\j=J\uppercase\expandafter{\expandafter
                        \takethree\topmark}}
\def\topsecno{\expandafter\takeone\topmark}
\def\takeone#1#2#3{#1}

\def\takethree#1#2#3{#3}
\def\nullsec{\eightrm\kern-2em} 

\let\page=\pagebody \raggedbottom
\def\normaloutput#1#2#3{\ifodd\pageno\hoffset=\pageshift\fi
 \shipout\vbox{
  \vbox to\fullpageheight{
  \iftitle\global\titlefalse
  \else\hbox to\pagewidth{\vbox to10pt{}\ifodd\pageno #3\else#2\fi}\fi
  \vfill#1}} 
  \global\advance\pageno by1}

\gtitle={} 
\mark{\noexpand\nullsec0{\the\gtitle}}
\def\title{\expandafter\uppercase\expandafter{\jobname}}
\def\topofcontents{\centerline{\titlefont\title}\vskip.7in
  \vfill} 
\def\botofcontents{\vfill
  \centerline{\covernote}} 
\def\covernote{}
\def\contentspagenumber{0} 
\newdimen\pagewidth \pagewidth=6.5in 
\newdimen\pageheight \pageheight=8.7in 
\newdimen\fullpageheight \fullpageheight=9in 
\newdimen\pageshift \pageshift=0in 

\def\setpage{\hsize\pagewidth\vsize\pageheight} 
\def\contentsfile{\jobname.toc} 
\def\readcontents{\input \contentsfile}
\def\readindex{\input \jobname.idx}
\def\readsections{\input \jobname.scn}

\newwrite\cont
\output{\setbox0=\page 
  \openout\cont=\contentsfile
       \write\cont{\catcode `\noexpand\@=11\relax}   
  \global\output{\normaloutput\page\lheader\rheader}}
\setpage
\vbox to \vsize{} 

\def\ch{\note{The following sections were changed by the change file:}
  \let\*=\relax}
\newbox\sbox 
\newbox\lbox 
\def\inx{\par\vskip6pt plus 1fil 
  \def\page{\box255 } \normalbottom
  \write\cont{} 
       \write\cont{\catcode `\noexpand\@=12\relax}   
  \closeout\cont 
  \output{\ifpagesaved\normaloutput{\box\sbox}\lheader\rheader\fi
    \global\setbox\sbox=\page \global\pagesavedtrue}
  \pagesavedfalse \eject 
  \setbox\sbox\vbox{\unvbox\sbox} 
  \vsize=\pageheight \advance\vsize by -\ht\sbox 
  \hsize=.5\pagewidth \advance\hsize by -10pt
  \parfillskip 0pt plus .6\hsize 
  \def\lr{L} 
  \output{\if L\lr\global\setbox\lbox=\page \gdef\lr{R}
    \else\normaloutput{\vbox to\pageheight{\box\sbox\vss
        \hbox to\pagewidth{\box\lbox\hfil\page}}}\lheader\rheader
    \global\vsize\pageheight\gdef\lr{L}\global\pagesavedfalse\fi}
  \message{Index:}
  \parskip 0pt plus .5pt
  \outer\def\I##1, {\par\hangindent2em\noindent##1:\kern1em} 
  \def\[##1]{$\underline{##1}$} 
  \rm \rightskip0pt plus 2.5em \tolerance 10000 \let\*=\lapstar
  \hyphenpenalty 10000 \parindent0pt
  \readindex}
\def\fin{\par\vfill\eject 
  \ifpagesaved\null\vfill\eject\fi 
  \if L\lr\else\null\vfill\eject\fi 
  \parfillskip 0pt plus 1fil
  \def\grouptitle{NAMES OF THE SECTIONS}
  \let\topsecno=\nullsec
  \message{Section names:}
  \output={\normaloutput\page\lheader\rheader}
  \setpage
  \def\note##1##2.{\quad{\eightrm##1~##2.}}
  \def\Q{\note{Cited in section}} 
  \def\Qs{\note{Cited in sections}} 
  \def\U{\note{Used in section}} 
  \def\Us{\note{Used in sections}} 
  \def\I{\par\hangindent 2em}\let\*=*
  \readsections}
\def\con{\par\vfill\eject 
  \rightskip 0pt \hyphenpenalty 50 \tolerance 200
  \setpage \output={\normaloutput\page\lheader\rheader}
  \titletrue 
  \pageno=\contentspagenumber
  \def\grouptitle{TABLE OF CONTENTS}
  \message{Table of contents:}
  \topofcontents
  \line{\hfil Section\hbox to3em{\hss Page}}
  \let\ZZ=\contentsline
  \readcontents\relax 
  \botofcontents \end} 
\def\contentsline#1#2#3#4{\ifnum#2=0 \smallbreak\fi
    \line{\consetup{#2}#1
      \rm\leaders\hbox to .5em{.\hfil}\hfil\ #3\hbox to3em{\hss#4}}}
\def\consetup#1{\ifcase#1 \bf 
  \or 
  \or \hskip2em 
  \or \hskip4em 
  \or \hskip6em 
  \or \hskip8em 
  \or \hskip10em 
  \else \hskip12em \fi} 
\def\noinx{\let\inx=\end} 
\def\nosecs{\let\FIN=\fin \def\fin{\let\parfillskip=\end \FIN}}
\def\nocon{\let\con=\end} 
\def\today{\ifcase\month\or
  January\or February\or March\or April\or May\or June\or
  July\or August\or September\or October\or November\or December\fi
  \space\number\day, \number\year}
\newcount\twodigits
\def\hours{\twodigits=\time \divide\twodigits by 60 \printtwodigits
  \multiply\twodigits by-60 \advance\twodigits by\time :\printtwodigits}
\def\gobbleone1{}
\def\printtwodigits{\advance\twodigits100
  \expandafter\gobbleone\number\twodigits
  \advance\twodigits-100 }
\def\TeX{{\ifmmode\it\fi
   \leavevmode\hbox{T\kern-.1667em\lower.424ex\hbox{E}\hskip-.125em X}}}
\def\,{\relax\ifmmode\mskip\thinmuskip\else\thinspace\fi}
\def\datethis{\def\startsection{\leftline{\sc\today\ at \hours}\bigskip
  \let\startsection=\stsec\stsec}}

\nocon 
\font\Large=cmr12

\def\ref#1{\hbox{(see [#1])}}


\centerline{ }
\vskip2cm

\centerline{\Large The DIR Net}
\centerline{\Large A Distributed System for Detection, Isolation, and Recovery}
\vskip0.5cm
\centerline{Vincenzo De Florio}
\centerline{Katholieke Universiteit Leuven}
\centerline{Departement Elektrotechniek}
\centerline{Afdeling ESAT/ACCA}
\vskip0.8cm
\centerline{\sc Technical Report ESAT/ACCA/1998/1}
\centerline{(pre-release TEX/0.6)}
\vskip12pt
\centerline{7 May, 1998}
\centerline{(revised on 27 April, 2015)}

\vfill\eject

\N{1}{1}The DIR net.

This document describes the DIR net [1], a distributed environment
which is part of the EFTOS fault tolerance framework [2].
The DIR net is a system consisting of two components, called
DIR Manager (or, shortly, the manager) and DIR Backup Agent
(shortly, the backup). One manager and a set of backups is
located in the system to be `guarded', one component per node.
At this point the DIR net weaves a web which substantially
does two things:

\item{$\bullet$} makes itself tolerant to a number of possible faults, and
\item{$\bullet$} gathers information pertaining the run of the user
application.

As soon as an error occurs {\it within the DIR net\/}, the system
executes built-in recovery actions that allow itself to continue
processing despite a number of hardware/software faults, possibly
doing a graceful degradation of its features; when an error occurs
{\it in the user application\/}, the DIR net, by means of custom-
and user-defined detection tools, is informed of such events and
runs one or more recovery strategies, both built-in and
coded by the user using an ancillary compile-time tool, the
{\sc rl} translator [3]. Such tools translates the user-defined
strategies into a binary ``R-code'', i.e., a pseudo-code interpreted
by a special component of the DIR net, the Recovery Interpreter,
{\sc rint} (in a sense, {\sc rint} is a r-code virtual machine.)

This document describes the generic component of the DIR net,
a function which can behave either as manager or as backup.

\fi

\M{2}The first mailbox-id available to the DIR-net
\Y\B\4\D$\.{DIR\_MBOX\_OFFSET}$ \5
\T{20}\par
\fi

\M{3}Same as above, but for Alias-id's
\Y\B\4\D$\.{DIR\_ALIAS\_OFFSET}$ \5
\T{20}\par
\fi

\M{4}On each node $n\in [0, \PB{\.{MAX\_PROCS}}[$ a DIR net component is to
run.
This component can be addresses internally by means of mailbox $\PB{%
\.{MBOX}}(n)$
and externally (from other nodes) by means of alias $\PB{\.{ALIAS}}(n)$.

\Y\B\4\D$\.{MBOX}(\|i)$ \5
\.{DIR\_MBOX\_OFFSET}\par
\B\4\D$\.{IAT\_MBOX}$ \5
$\.{DIR\_MBOX\_OFFSET}+{}$\T{1}\par
\B\4\D$\.{RINT\_MBOX}$ \5
$\.{DIR\_MBOX\_OFFSET}+{}$\T{2}\par
\B\4\D$\.{DB\_MBOX}$ \5
$\.{DIR\_MBOX\_OFFSET}+{}$\T{3}\par
\B\4\D$\.{TOM\_MBOX}$ \5
$\.{DIR\_MBOX\_OFFSET}+{}$\T{4}\par
\B\4\D$\.{ALIAS}(\|i)$ \5
\.{DIR\_ALIAS\_OFFSET}\par
\B\4\D$\.{IAT\_ALIAS}$ \5
\.{IAT\_MBOX}\par
\B\4\D$\.{IA\_FLAG\_TIMEOUT}$ \5
\T{10}\par
\B\4\D$\.{IA\_FLAG\_CYCLIC}$ \5
\.{TOM\_CYCLIC}\par
\B\4\D$\.{IA\_FLAG\_DEADLINE}$ \5
\.{IMALIVE\_CLEAR\_TIMEOUT}\par
\B\4\D$\.{MIA\_TIMEOUT}$ \5
\T{15}\par
\B\4\D$\.{MIA\_CYCLIC}$ \5
\.{TOM\_CYCLIC}\par
\B\4\D$\.{MIA\_DEADLINE}$ \5
\.{MIA\_SEND\_TIMEOUT}\par
\B\4\D$\.{TAIA\_TIMEOUT}$ \5
\T{20}\par
\B\4\D$\.{TAIA\_CYCLIC}$ \5
\.{TOM\_CYCLIC}\par
\B\4\D$\.{TAIA\_DEADLINE}$ \5
\.{TAIA\_RECV\_TIMEOUT}\par
\B\4\D$\.{TEIF\_TIMEOUT}$ \5
\T{30}\par
\B\4\D$\.{TEIF\_CYCLIC}$ \5
\.{TOM\_NON\_CYCLIC}\par
\B\4\D$\.{TEIF\_DEADLINE}$ \5
\.{IMALIVE\_SET\_TIMEOUT}\par
\B\4\D$\.{IA\_FLAG\_TIMEOUT\_B}$ \5
\T{50}\par
\B\4\D$\.{IA\_FLAG\_CYCLIC\_B}$ \5
\.{TOM\_CYCLIC}\par
\B\4\D$\.{IA\_FLAG\_DEADLINE\_B}$ \5
\.{IMALIVE\_CLEAR\_TIMEOUT}\par
\B\4\D$\.{MIA\_TIMEOUT\_B}$ \5
\T{55}\par
\B\4\D$\.{MIA\_CYCLIC\_B}$ \5
\.{TOM\_CYCLIC}\par
\B\4\D$\.{MIA\_DEADLINE\_B}$ \5
\.{MIA\_RECV\_TIMEOUT}\par
\B\4\D$\.{TAIA\_TIMEOUT\_B}$ \5
\T{60}\par
\B\4\D$\.{TAIA\_CYCLIC\_B}$ \5
\.{TOM\_CYCLIC}\par
\B\4\D$\.{TAIA\_DEADLINE\_B}$ \5
\.{TAIA\_SEND\_TIMEOUT}\par
\B\4\D$\.{TEIF\_TIMEOUT\_B}$ \5
\T{70}\par
\B\4\D$\.{TEIF\_CYCLIC\_B}$ \5
\.{TOM\_NON\_CYCLIC}\par
\B\4\D$\.{TEIF\_DEADLINE\_B}$ \5
\.{IMALIVE\_SET\_TIMEOUT}\par
\B\4\D$\.{IAT\_TIMEOUT}$ \5
\T{40}\par
\B\4\D$\.{IAT\_CYCLIC}$ \5
\.{TOM\_CYCLIC}\par
\B\4\D$\.{IAT\_DEADLINE}$ \5
\.{IMALIVE\_SET\_TIMEOUT}\par
\B\4\D$\.{INJECT\_FAULT\_TIMEOUT}$ \5
\T{6}\par
\B\4\D$\.{INJECT\_FAULT\_DEADLINE}$ \5
\T{6000000}\C{ 6 secs }\par
\Y\B\X5:Global Variables and \PB{$\#$ \&{include}}'s\X\7
\X6:Generic component of the DIR net\X\6
\X52:Alarm function\X\6
\X61:I'm Alive Task\X\6
\X56:\PB{\\{GetState}} and \PB{\\{SetState}}\X\6
\X59:TEX routines simulated on EPX\X\6
\X60:DIR Print Message\X\par
\fi

\M{5}We need to include a number of header files, i.e.,
those pertaining the timeout manager, ...

\Y\B\4\X5:Global Variables and \PB{$\#$ \&{include}}'s\X${}\E{}$\6
\8\#\&{include} \.{<stdio.h>}\6
\8\#\&{include} \.{<stdlib.h>}\6
\8\#\&{include} \.{<sys/root.h>}\6
\8\#\&{include} \.{<sys/logerror.h>}\6
\8\#\&{include} \.{<sys/link.h>}\6
\8\#\&{include} \.{<sys/select.h>}\6
\8\#\&{include} \.{<sys/time.h>}\6
\8\#\&{include} \.{<sys/thread.h>}\6
\8\#\&{include} \.{<sys/sem.h>}\6
\8\#\&{include} \.{<string.h>}\6
\8\#\&{include} \.{"tom.h"}\6
\8\#\&{include} \.{"timeouts.h"}\6
\8\#\&{include} \.{"dirdefs.h"}\6
\8\#\&{ifdef} \.{EPX1\_2}\6
\&{typedef} \&{int} \&{IDF};\6
\8\#\&{endif}\6
\8\#\&{include} \.{"dirtypes.h"}\6
\8\#\&{include} \.{"rcode.h"}\6
\8\#\&{include} \.{"trl.h"}\C{ TEX-specific includes {\it \&\/} types }\6
\8\#\&{ifdef} \.{TEX}\6
\8\#\&{include} \.{<links.h>}\6
\8\#\&{include} \.{<remote\_mbox.h>}\6
\8\#\&{include} \.{<thread.h>}\6
\8\#\&{else}\6
\&{int} \\{TEXFirstActivation}${}\K\T{1};{}$\6
\&{typedef} \&{int} \&{Alias\_t};\6
\&{typedef} \&{int} \&{IDF};\6
\8\#\&{define} \.{MSG\_OK} \5\T{0}\6
\8\#\&{define} \.{INFINITE} \5\T{0}\6
\8\#\&{endif}\6
\8\#\&{ifdef} \.{EPX1\_2}\6
\8\#\&{include} \.{<sys/rrouter.h>}\6
\&{int} ${}\\{RemoteSendMessage}(\&{int},\39\&{int},\39{}$\&{char} ${}{*},\39%
\&{int});{}$\6
\&{int} ${}\\{TEXSendMessage}(\&{int},\39{}$\&{char} ${}{*},\39\&{int});{}$\6
\&{int} ${}\\{TEXReceiveMessage}(\&{int},\39{}$\&{char} ${}{*},\39{}$\&{int}
${}{*},\39\&{int});{}$\6
\&{int} \\{GetRoot}(\&{void});\6
\8\#\&{endif}\6
\&{int}  \\{TEXGetState} ( \\{status\_t} $*$ )  ; \&{int}  \\{TEXSetState} ( %
\\{status\_t} $*$ )  ;\6
\8\#\&{ifndef} \.{EPX}\7
\&{int} ${}\\{Export}(\&{Alias\_t},\39\&{IDF});{}$\7
\.{STATUS}\\{TEXGetTaskStatus}(\&{IDF});\6
\8\#\&{endif}\7
\&{int} \\{TEXGetNumTasks}(\&{void});\6
\&{int} \\{TEXStopTask}(\&{void});\6
\&{int} \\{TEXRestartTask}(\&{void});\6
\&{int} \\{flag}${}\K\T{0};{}$\6
\&{char} ${}{*}\\{role2ascii}(\&{int});{}$\6
\&{char} ${}{*}\\{DIRPrintTimeout}(\&{int});{}$\6
\&{char} ${}{*}\\{DIRPrintMessage}(\&{int});{}$\6
\&{char} ${}{*}\\{DIRPrintCode}(\&{int});$ \&{int}  \\{send\_timeout\_message}
( \.{TOM} $*$ )  ;\7
\&{extern} \\{Semaphore\_t}\\{sem}${},{}$ ${}{*}\\{tom\_sem};{}$\6
\8\#\&{ifndef} \.{\_TOM\_\_H\_}\6
\&{typedef} \&{struct} ${}\{{}$\1\6
\&{int} \\{running};\6
\&{int} \\{deadline};\7
${}\&{int}({*}\\{alarm}){}$(\&{struct} \.{TOM} ${}{*});{}$\7
\&{unsigned} \&{char} \\{id}${},{}$ \\{subid};\6
\&{unsigned} \&{char} \\{cyclic};\6
\&{unsigned} \&{char} \\{suspended};\2\6
${}\}{}$ \&{timeout\_t};\6
\&{typedef} \&{struct} \&{block\_t} ${}\{{}$\1\6
\&{struct} \&{block\_t} ${}{*}\\{next};{}$\6
\&{timeout\_t} \\{timeout};\6
\&{unsigned} \&{char} \\{used};\2\6
${}\}{}$ \&{block\_t};\6
\&{typedef} \&{struct} \&{TOM} ${}\{{}$\1\6
\&{block\_t} ${}{*}\\{top};{}$\6
\&{int} \\{tom\_id};\6
\&{block\_t} ${}{*}\\{block\_stack};{}$\6
\&{int} \\{block\_sp};\7
${}\&{int}({*}\\{default\_alarm}){}$(\&{struct} \&{TOM} ${}{*});{}$\6
${}\\{LinkCB\_t}*\\{link}[\T{2}];{}$\7
\&{unsigned} \&{int} \\{starting\_time};\2\6
${}\}{}$ \&{TOM};\6
\&{typedef} \&{struct} ${}\{{}$\1\6
\&{timeout\_t} ${}{*}\\{timeout};{}$\6
\&{unsigned} \&{char} \\{code};\2\6
${}\}{}$ \&{tom\_message\_t};\6
\8\#\&{endif}\6
\8\#\&{ifndef} \.{\_\_DIR\_TYPES\_}\6
\&{typedef} \&{struct} ${}\{{}$\1\6
\&{char} \\{configuration};\6
\&{int} \\{runlevel};\2\6
${}\}{}$ \&{DIR\_state\_t};\6
\&{typedef} \&{struct} ${}\{{}$\1\6
\&{char} \\{primary};\6
\&{char} \\{role};\2\6
${}\}{}$ \&{status\_t};\6
\8\#\&{endif}\6
\8\#\&{define} \.{MAXARG} \5\T{5}\6
\&{typedef} \&{struct} ${}\{{}$\1\6
\&{int} \\{arg}[\.{MAXARG}];\6
\&{int} \\{subid};\6
\&{int} \\{type};\6
\&{char} \\{local};\2\6
${}\}{}$ \&{message\_t};\6
\&{int} \\{IA\_flag};\6
\&{message\_t} \\{message};\6
\&{extern} \\{DIR\_db\_t}\\{db};\6
\&{int} \\{errors};\par
\U4.\fi

\M{6}On every node a \PB{\\{DIRNetGenericComponent}(\,)} function is run. This
function first runs a protocol to understand its role
and who is the manager, and to build or re-build a global database;
after this phase it runs either as a manager or as a
backup agent.

Merging the codes for the backup and the manager into one
running component has two major benefits: {\it (1)\/} you need to foresee
$n$ replicas of this generic component, without specifying
which role each replica has to play, and {\it (2)\/} a lot of
code is shared between the two components.

\Y\B\4\X6:Generic component of the DIR net\X${}\E{}$\6
\&{void} \\{DIRNetGenericComponent}(\&{void})\1\1\2\2\6
${}\{{}$\1\6
\X7:Variables local to the Manager and the Backup\X\6
${}\\{InitSem}(\\{tom\_sem},\39\T{1});{}$\6
\X8:DIR net initialisation\X\6
\X69:Spawn the I'm Alive Task\X\6
\&{if} ${}(\\{role}\E\.{DIR\_MANAGER}){}$\5
${}\{{}$\1\6
\X9:DIR net manager\X\6
\4${}\}{}$\2\6
\&{else}\5
${}\{{}$\1\6
\X33:DIR net backup agent\X\6
\4${}\}{}$\2\6
\\{TEXStopTask}(\,);\6
\4${}\}{}$\2\par
\U4.\fi

\M{7}These variables are shared between the Manager and
the Backup Agent.
\Y\B\4\X7:Variables local to the Manager and the Backup\X${}\E{}$\6
\&{status\_t} \\{mystate};\7
\.{STATUS}\\{status};\7
\&{int} \\{role}${},{}$ \\{managerid};\6
\&{TOM} ${}{*}\\{tom};{}$\6
\&{timeout\_t} \\{mia}[\.{MAX\_PROCS}]${},{}$ \\{taia}[\.{MAX\_PROCS}]${},{}$ %
\\{teif}${},{}$ \\{ia}${},{}$ \\{inject};\6
\&{char} \\{suspicion\_period}[\.{MAX\_PROCS}];\6
\&{int} \|n${},{}$ \\{sender};\par
\U6.\fi

\M{8}This function initialises the DIR net. It checks whether this node
and this components are both ``new'' (a node is new if it has never been
rebooted; a component is ``new'' if it is the original component of
this node, i.e., a primary). If so, it gets the role from the state,
as stored in the database, otherwise, it asks the neighbouring
nodes who is the manager. This section also takes care of building
or receiving the system database: if the node is new, then
a local database is built and the global database is created by
a broadcast session; otherwise, the global database is requested
from a remote node.
\Y\B\4\X8:DIR net initialisation\X${}\E{}$\6
$\\{TEXGetState}({\AND}\\{mystate});{}$\6
\&{if} ${}(\\{TEXFirstActivation}\W\\{mystate}.\\{primary}){}$\5
${}\{{}$\C{ only two roles are possible -- backup or manager }\1\6
\X57:Read your role from the RL script\X;\6
\&{if} ${}(\\{db}.\\{role}\E\.{DIR\_BACKUP}){}$\1\5
${}\\{role}\K\.{DIR\_BACKUP};{}$\2\6
\&{else}\1\5
${}\\{role}\K\.{DIR\_MANAGER};{}$\2\6
${}\\{LogError}(\.{EC\_ERROR},\39\.{"Generic"},\39\.{"role\ of\ \%d\ is\ \%s"},%
\39\\{GetRoot}(\,),\39\\{role2ascii}(\\{role}));{}$\6
\\{fork}(\,);\C{ export your main mailbox }\6
${}\\{Export}(\.{ALIAS}(\\{GetRoot}(\,)),\39\.{MBOX}(\\{GetRoot}(\,))){}$;\C{
export your I'm Alive Task's mailbox }\6
\\{Export}((\&{Alias\_t}) \.{IAT\_ALIAS}${},\39{}$(\&{IDF}) \.{IAT\_MBOX});\C{
export the database mailbox }\6
\\{Export}((\&{Alias\_t}) \.{DB\_MBOX}${},\39{}$(\&{IDF}) \.{DB\_MBOX});\6
\X58:Check who is the manager according to the RL script\X;\6
\4${}\}{}$\2\6
\&{else}\5
${}\{{}$\1\6
\X54:send \PB{\.{WITM}} to all\X\6
\X55:wait for \PB{\.{NMI}} messages to come\X\6
\&{if} ${}(\\{message}.\\{arg}[\T{0}]\E\\{GetRoot}(\,)){}$\1\5
${}\\{role}\K\.{DIR\_MANAGER};{}$\2\6
\&{else}\1\5
${}\\{role}\K\.{DIR\_BACKUP};{}$\2\6
${}\\{managerid}\K\\{message}.\\{arg}[\T{0}];{}$\6
\4${}\}{}$\2\6
\&{if} (\\{TEXFirstActivation})\5
${}\{{}$\1\6
\&{int} \|i;\7
${}\|n\K\\{TEXGetNumTasks}(\,);{}$\6
\X44:store number of tasks (\PB{\|n});\X\6
\&{for} ${}(\|i\K\T{0};{}$ ${}\|i<\|n;{}$ ${}\|i\PP){}$\5
${}\{{}$\1\6
${}\\{status}\K\\{TEXGetTaskStatus}{}$((\&{IDF}) \|i);\6
\X45:store in local database(i, status)\X\6
\4${}\}{}$\2\6
\X46:broadcast local database and receive the others' databases\X\6
\X49:build a global database\X\6
\X50:mark as `reboot-resistant' the whole database via \PB{\\{DataReset}(\,)}\X%
\6
\4${}\}{}$\2\6
\&{else}\5
${}\{{}$\1\6
\X51:request a copy of the global database\X\6
\4${}\}{}$\2\6
${}\\{LogError}(\.{EC\_MESS},\39\.{"Generic"},\39\.{"global\ database\
OK"}){}$;\par
\U6.\fi

\M{9}The code for the manager of the DIR net.
\Y\B\4\X9:DIR net manager\X${}\E{}$\6
${}\{{}$\1\6
${}\\{managerid}\K\\{GetRoot}(\,);{}$\6
${}\\{LogError}(\.{EC\_ERROR},\39\.{"Manager"},\39\.{"Manager\ starts..."}){}$;%
\C{ the alarm of these timeouts simply sends a message of id 	
$<$timeout-type$>$, subid = \PB{\\{subid}} to the manager 	 }\6
\X10:insert timeouts (IA-flag-timeout, MIA-timeouts, TAIA-timeouts)\X\6
\X14:clear the \PB{\\{suspicion\_period}[\,]}'s\X\6
\X16:clear IA-flag\X\6
${}\\{LogError}(\.{EC\_ERROR},\39\.{"Manager"},\39\.{"Manager\ activates\ I}\)%
\.{AT..."});{}$\6
\X17:activate IAT\X\6
${}\\{LogError}(\.{EC\_ERROR},\39\.{"Manager"},\39\.{"Manager\ loop\ starts}\)%
\.{..."});{}$\6
\X18:manager loop (waiting for incoming messages)\X\6
\4${}\}{}$\C{ end manager }\2\par
\U6.\fi

\M{10}The manager initialises a set of timeout objects and inserts
them into the timeout list [4] [5].
\Y\B\4\X10:insert timeouts (IA-flag-timeout, MIA-timeouts, TAIA-timeouts)\X${}%
\E{}$\6
${}\{{}$\1\6
${}\\{tom}\K\\{tom\_init}(\\{send\_timeout\_message});{}$\6
\X11:declare and insert MIA timeouts\X\6
\X12:declare and insert the IA-flag timeout\X\6
\X13:declare and insert TAIA timeouts\X\6
\4${}\}{}$\2\par
\U9.\fi

\M{11}At most every \PB{\.{MIA\_DEADLINE}} ticks a ``Manager Is Alive'' (\PB{%
\.{MIA}})
message needs to start towards a backup.
\Y\B\4\X11:declare and insert MIA timeouts\X${}\E{}$\6
${}\{{}$\1\6
\&{int} \|i;\7
\&{for} ${}(\|i\K\T{0};{}$ ${}\|i<\.{MAX\_PROCS};{}$ ${}\|i\PP){}$\5
${}\{{}$\1\6
\&{if} ${}(\|i\E\\{GetRoot}(\,)){}$\1\5
\&{continue};\2\6
${}\\{tom\_declare}(\\{mia}+\|i,\39\.{MIA\_CYCLIC},\39\.{TOM\_SET\_ENABLE},\39%
\.{MIA\_TIMEOUT},\39\|i,\39\.{MIA\_DEADLINE});{}$\6
${}\\{tom\_insert}(\\{tom},\39\\{mia}+\|i);{}$\6
\4${}\}{}$\2\6
\8\#\&{ifdef} \.{INJECT}\6
${}\\{tom\_declare}({\AND}\\{inject},\39\.{TOM\_NON\_CYCLIC},\39\.{TOM\_SET%
\_ENABLE},\39\.{INJECT\_FAULT\_TIMEOUT},\39\|i,\39\.{INJECT\_FAULT%
\_DEADLINE});{}$\6
${}\\{tom\_insert}(\\{tom},\39{\AND}\\{inject});{}$\6
\8\#\&{endif}\6
\4${}\}{}$\2\par
\U10.\fi

\M{12}Every \PB{\.{IMALIVE\_CLEAR\_TIMEOUT}} ticks at most the IA-flag must
be cleared, or the IAT will consider this component as crashed.
This is accomplished by means of the following cyclic timeout:
\Y\B\4\X12:declare and insert the IA-flag timeout\X${}\E{}$\6
${}\{{}$\1\6
${}\\{tom\_declare}({\AND}\\{ia},\39\.{IA\_FLAG\_CYCLIC},\39\.{TOM\_SET%
\_ENABLE},\39\.{IA\_FLAG\_TIMEOUT},\39\T{1},\39\.{IMALIVE\_CLEAR\_TIMEOUT});{}$%
\6
${}\\{tom\_insert}(\\{tom},\39{\AND}\\{ia});{}$\6
\4${}\}{}$\2\par
\Us10\ET34.\fi

\M{13}At most every \PB{\.{TAIA\_DEADLINE}} ticks a ``This Agent Is Alive''
(\PB{\.{TAIA}}) message coming from a backup needs to be received
by the Manager.
\Y\B\4\X13:declare and insert TAIA timeouts\X${}\E{}$\6
${}\{{}$\1\6
\&{int} \|i;\7
\&{for} ${}(\|i\K\T{0};{}$ ${}\|i<\.{MAX\_PROCS};{}$ ${}\|i\PP){}$\5
${}\{{}$\1\6
\&{if} ${}(\|i\E\\{GetRoot}(\,)){}$\1\5
\&{continue};\2\6
${}\\{tom\_declare}(\\{taia}+\|i,\39\.{TAIA\_CYCLIC},\39\.{TOM\_SET\_ENABLE},%
\39\.{TAIA\_TIMEOUT},\39\|i,\39\.{TAIA\_DEADLINE});{}$\6
${}\\{tom\_insert}(\\{tom},\39\\{taia}+\|i);{}$\6
\4${}\}{}$\2\6
\4${}\}{}$\2\par
\U10.\fi

\M{14}Suspicion periods are globally set to off
\Y\B\4\X14:clear the \PB{\\{suspicion\_period}[\,]}'s\X${}\E{}$\6
${}\{{}$\1\6
\&{int} \|i;\7
\&{for} ${}(\|i\K\T{0};{}$ ${}\|i<\.{MAX\_PROCS};{}$ ${}\|i\PP){}$\5
${}\{{}$\1\6
${}\\{suspicion\_period}[\|i]\K\T{0};{}$\6
\4${}\}{}$\2\6
\4${}\}{}$\2\par
\U9.\fi

\M{15}This is the same as above, but for the backup agent. It
only manages one suspicion period, {\it viz.\/} the one of the
manager; let us choose 0 as the manager's suspicion period.
\Y\B\4\X15:clear \PB{\\{suspicion\_period}}\X${}\E{}$\6
${*}\\{suspicion\_period}\K\T{0}{}$;\par
\U33.\fi

\M{16}\PB{\\{IA\_flag}} is shared between this component and its I'm Alive
Task. At initialisation time and at each new message of type
\PB{\.{IA\_FLAG\_TIMEOUT}} this flag has to be cleared.
\Y\B\4\X16:clear IA-flag\X${}\E{}$\6
$\\{IA\_flag}\K\T{0}{}$;\par
\Us9, 18, 33\ETs37.\fi

\M{17}This message wakes up the I'm Alive Task, which will
start checking periodically whether the IA-flag has been
cleared. (Note: each component knows that there are
nProcs-1 fellows in the dirnet. It is assumed that
fellow $i$ gets mail on mailbox $i$, so the first
\PB{\\{nProcs}}-1 integers are reserved for this reason.
Furthermore, each node shall export its mailbox via
\PB{\\{Export}((\&{Alias\_t}) \\{GetRoot}(\,)${},{}$(\&{IDF}) \\{GetRoot}(%
\,))}.)
Mailbox whose id and Alias is \PB{\\{nProcs}} is assumed to be
the I'm Alive Task, mailbox \PB{\\{nProcs}}+1 is the recovery thread.

\Y\B\4\X17:activate IAT\X${}\E{}$\6
$\\{message}.\\{type}\K\.{ROUSE};{}$\6
${}\\{message}.\\{subid}\K\\{GetRoot}(\,);{}$\6
${}\\{TEXSendMessage}(\.{IAT\_MBOX},\39{}$(\&{char} ${}{*}){}$ ${}{\AND}%
\\{message},\39{}$\&{sizeof} (\\{message}));\par
\Us9, 18, 33\ETs37.\fi

\M{18}This loop is the real core of the manager. It has to deal
with a number of messages coming from the timeout manager,
its fellow backups, the recovery thread, the remote I'm Alive
Tasks. The core of the fault-tolerant strategy of the
DIR net is in here.
\Y\B\4\X18:manager loop (waiting for incoming messages)\X${}\E{}$\6
\&{while} (\T{1})\5
${}\{{}$\1\6
\X53:wait for an incoming message\X\6
\\{tom\_dump}(\\{tom});\6
\&{switch} ${}(\\{message}.\\{type}){}$\5
${}\{{}$\1\6
\4\&{case} \.{INJECT\_FAULT\_TIMEOUT}:\6
${}\\{LogError}(\.{EC\_ERROR},\39\.{"Manager\ loop"},\39\.{"Fault\
injection"});{}$\6
\\{tom\_close}(\\{tom});\C{ the time-out manager is detached }\6
\&{break};\6
\4\&{case} \.{IA\_FLAG\_TIMEOUT}:\6
${}\\{LogError}(\.{EC\_ERROR},\39\.{"Manager\ loop"},\39\.{"IA\_FLAG\_TIMEOUT\
mes}\)\.{sage\ ->\ clear\ IA-fla}\)\.{g."}){}$;\C{ time to clear the IA-flag! }%
\6
\X16:clear IA-flag\X\6
\&{break};\6
\4\&{case} \.{MIA\_TIMEOUT}:\6
${}\\{LogError}(\.{EC\_ERROR},\39\.{"Manager\ loop"},\39\.{"MIA\_TIMEOUT\
message}\)\.{\ (time\ to\ send\ a\ MIA}\)\.{\ to\ Backup\ \%d)."},\39%
\\{message}.\\{subid}){}$;\C{ time to send a \PB{\.{MIA}} to a backup }\6
\X19:send \PB{\.{MIA}} to backup \PB{\\{subid}}\X\6
${}\\{tom\_dump}(\\{tom}+\\{message}.\\{subid});{}$\6
${}\\{tom\_renew}(\\{tom},\39\\{mia}+\\{message}.\\{subid});{}$\6
\&{break};\6
\4\&{case} \.{DB}:\6
${}\\{LogError}(\.{EC\_ERROR},\39\.{"Manager\ loop"},\39\.{"DB\ message."}){}$;%
\C{ a message which modifies the db }\6
\&{if} ${}(\\{message}.\\{subid}\E\\{GetRoot}(\,){}$)\C{ the message is local }%
\6
${}\{{}$\1\6
\X20:renew \PB{\.{MIA}} timeout on all \PB{\\{subid}}'s\X\6
\X22:update your copy of the db\X\6
\X21:broadcast modifications to db\X\6
\&{break};\6
\4${}\}{}$\2\6
\&{else}\C{ if it's a remote message, it's also a piggybacked \PB{\.{TAIA}} }\6
${}\{{}$\1\6
\X22:update your copy of the db\X\C{ don't break }\6
\4${}\}{}$\C{ don't break }\2\6
\4\&{case} \.{TAIA}:\6
${}\\{LogError}(\.{EC\_ERROR},\39\.{"Manager\ loop"},\39\.{"TAIA\ message\ from%
\ n}\)\.{ode\ \%d."},\39\\{message}.\\{subid}){}$;\C{ if a \PB{\.{TAIA}} comes
in or a remote \PB{\.{DB}} message comes in... }\6
\&{if} ${}(\R\\{tom\_ispresent}(\\{tom},\39\\{taia}+\\{message}.\\{subid})){}$\5
${}\{{}$\1\6
\X23:insert TAIA-timeout, subid=\PB{\\{subid}}\X\6
\X24:broadcast \PB{\.{NIUA}}\X\C{ node is up again! }\6
\4${}\}{}$\C{ if you get a \PB{\.{TAIA}} while expecting a \PB{\.{TEIF}}, then
		   no problem, simply get out of the suspicion period 		 }\2\6
\&{if} ${}(\\{suspicion\_period}[\\{message}.\\{subid}]\E\.{TRUE}){}$\5
${}\{{}$\1\6
${}\\{LogError}(\.{EC\_ERROR},\39\.{"Manager"},\39\.{"got\ a\ TAIA\ while\ ex}%
\)\.{pecting\ a\ TEIF\ =>\ ge}\)\.{t\ out\ of\ the\ suspici}\)\.{on\
period"});{}$\6
${}\\{suspicion\_period}[\\{message}.\\{subid}]\K\.{FALSE};{}$\6
\4${}\}{}$\2\6
\&{else}\5
${}\{{}$\1\6
${}\\{LogError}(\.{EC\_ERROR},\39\.{"Manager"},\39\.{"got\ a\ due\ TAIA\ in\ t}%
\)\.{ime\ --\ renewing\ TAIA}\)\.{\ timeout\ \%d."},\39\\{message}.%
\\{subid});{}$\6
\X25:renew TAIA-timeout, \PB{\\{subid}}\X\6
\4${}\}{}$\2\6
\&{break};\6
\4\&{case} \.{TAIA\_TIMEOUT}:\6
${}\\{LogError}(\.{EC\_ERROR},\39\.{"Manager\ loop"},\39\.{"TAIA\_TIMEOUT\
messag}\)\.{e:\ no\ heartbeat\ from}\)\.{\ Backup\ \%d\ --\ enteri}\)\.{ng\
suspicion\ period"},\39\\{message}.\\{subid}){}$;\C{ no heartbeat from a remote
component\dots{} enter 	   a suspicion period then. 	 }\6
${}\\{suspicion\_period}[\\{message}.\\{subid}]\K\.{TRUE};{}$\6
\X26:insert \PB{\.{TEIF}} timeout\X\6
\X27:delete timeout (TAIA-timeout, \PB{\\{subid}})\X\6
\&{break};\6
\4\&{case} \.{TEIF}:\6
${}\\{LogError}(\.{EC\_ERROR},\39\.{"Manager\ loop"},\39\.{"TEIF\ message:\
IAT@n}\)\.{ode\%d\ sent\ an\ alarm\ }\)\.{and\ went\ to\ sleep."},\39%
\\{message}.\\{subid}){}$;\C{ a \PB{\.{TEIF}} message has been sent from a IAT:
	   as its last action, the IAT went to sleep 	 }\6
\&{if} ${}(\\{suspicion\_period}[\\{message}.\\{subid}]\E\.{TRUE}){}$\5
${}\{{}$\C{ 			$<$delete timeout (TAIA-timeout, \PB{\\{subid}})$>$ 			 }\1\6
\X28:delete timeout (TEIF-timeout, \PB{\\{subid}})\X\6
${}\\{suspicion\_period}[\\{message}.\\{subid}]\K\.{FALSE}{}$;\C{ agent
recovery will spawn a clone of the 			   backup \PB{\\{subid}}. If no backup
clones 			   are available, the entire node will 			   be rebooted. 			 }\6
\X29:Agent-Recovery(\PB{\\{subid}})\X\6
\4${}\}{}$\2\6
\&{else}\5
${}\{{}$\1\6
\&{if} ${}(\\{message}.\\{subid}\E\\{GetRoot}(\,)){}$\5
${}\{{}$\1\6
\X16:clear IA-flag\X\6
\X17:activate IAT\X\6
\4${}\}{}$\2\6
\&{else}\5
${}\{{}$\1\6
\&{int} \\{to}${}\K\\{message}.\\{subid};{}$\7
${}\\{message}.\\{type}\K\.{ENIA}{}$;\C{ i.e., ``ENable IAt'' }\6
${}\\{message}.\\{subid}\K\\{GetRoot}(\,);{}$\6
${}\\{RemoteSendMessage}(\\{to},\39\.{ALIAS}(\\{to}),\39{}$(\&{char} ${}{*}){}$
${}{\AND}\\{message},\39{}$\&{sizeof} (\\{message}));\6
\4${}\}{}$\2\6
\4${}\}{}$\2\6
\&{break};\6
\4\&{case} \.{TEIF\_TIMEOUT}:\6
${}\\{LogError}(\.{EC\_ERROR},\39\.{"Manager\ loop"},\39\.{"TEIF\_TIMEOUT\
messag}\)\.{e\ --\ the\ Manager\ con}\)\.{cludes\ that\ the\ susp}\)\.{ected\
node\ (\%d)\ has\ }\)\.{crashed."},\39\\{message}.\\{subid});{}$\6
\&{if} ${}(\\{suspicion\_period}[\\{message}.\\{subid}]\E\.{TRUE}){}$\5
${}\{{}$\1\6
\X27:delete timeout (TAIA-timeout, \PB{\\{subid}})\X\6
${}\\{suspicion\_period}[\\{message}.\\{subid}]\K\.{FALSE}{}$;\C{ the entire
node will be rebooted. 			 }\6
\X31:Node-Recovery(\PB{\\{subid}})\X\6
\4${}\}{}$\2\6
\&{break};\6
\4\&{case} \.{ENIA}:\6
${}\\{LogError}(\.{EC\_ERROR},\39\.{"Manager\ loop"},\39\.{"ENIA\
message."});{}$\6
\X16:clear IA-flag\X\C{ if an IAT gets an ``activate'' message while it's 		
active, that message is ignored }\6
\X17:activate IAT\X\6
\&{break};\6
\4\&{case} \.{WITM}:\6
${}\\{LogError}(\.{EC\_ERROR},\39\.{"Manager\ loop"},\39\.{"WITM\
message."});{}$\6
${}\\{sender}\K\\{message}.\\{subid};{}$\6
${}\\{message}.\\{subid}\K\\{message}.\\{arg}[\T{0}]\K\\{GetRoot}(\,);{}$\6
${}\\{message}.\\{type}\K\.{NMI};{}$\6
${}\\{RemoteSendMessage}(\\{sender},\39\.{ALIAS}(\\{sender}),\39{}$(\&{char}
${}{*}){}$ ${}{\AND}\\{message},\39{}$\&{sizeof} (\\{message}));\6
\&{break};\6
\4\&{case} \.{NIUA}:\6
${}\\{LogError}(\.{EC\_ERROR},\39\.{"Backup"},\39\.{"NIUA\ message\ --\ nod}\)%
\.{e\ \%d\ is\ up\ again\ --\ }\)\.{watching\ restarts...}\)\.{"},\39%
\\{message}.\\{subid}){}$;\C{ a Node Is Up Again: restart watchin' it }\6
\&{if} ${}(\R\\{tom\_ispresent}(\\{tom},\39\\{taia}+\\{message}.\\{subid})){}$%
\1\5
${}\\{tom\_insert}(\\{tom},\39\\{taia}+\\{message}.\\{subid});{}$\2\6
\&{break};\6
\4\&{case} \.{REQUEST\_DB}:\6
${}\\{LogError}(\.{EC\_ERROR},\39\.{"Backup"},\39\.{"REQUEST\_DB\ message.}\)%
\.{"}){}$;\C{ a node is requesting a full copy of the database }\6
${}\\{RemoteSendMessage}(\\{message}.\\{subid},\39\.{ALIAS}(\\{message}.%
\\{subid}),\39{}$(\&{char} ${}{*}){}$ ${}{\AND}\\{db},\39{}$\&{sizeof} (%
\\{db}));\6
\&{break};\6
\4\&{default}:\6
${}\\{LogError}(\.{EC\_ERROR},\39\.{"Backup"},\39\.{"Other\ messages."});{}$\6
\X32:deal with these other messages\X\6
\4${}\}{}$\C{ end switch (message.type) }\2\6
\X16:clear IA-flag\X\6
\X43:renew IA-flag-timeout\X\6
\4${}\}{}$\C{ end manager loop }\2\par
\U9.\fi

\M{19}A ``Manager Is Alive'' message needs to be sent to component
\PB{$\\{message}.\\{subid}$}:
\Y\B\4\X19:send \PB{\.{MIA}} to backup \PB{\\{subid}}\X${}\E{}$\6
$\\{message}.\\{type}\K\.{MIA}{}$;\C{ Note: MIA.arg[0] == managerid!! }\6
${}\\{message}.\\{arg}[\T{0}]\K\\{GetRoot}(\,);{}$\6
${}\\{RemoteSendMessage}(\\{message}.\\{subid},\39\.{ALIAS}(\\{message}.%
\\{subid}),\39{}$(\&{char} ${}{*}){}$ ${}{\AND}\\{message},\39{}$\&{sizeof} (%
\\{message}));\par
\U18.\fi

\M{20}A broadcast is about to take place---this implies that a suite
of implicit \PB{\.{MIA}}'s will be sent in piggybacking. As a consequence, all
\PB{\.{MIA\_SEND\_TIMEOUT}}'s needs to be renewed.
\Y\B\4\X20:renew \PB{\.{MIA}} timeout on all \PB{\\{subid}}'s\X${}\E{}$\6
${}\{{}$\1\6
\&{int} \|i;\7
\&{for} ${}(\|i\K\T{0};{}$ ${}\|i<\\{GetRoot}(\,);{}$ ${}\|i\PP){}$\5
${}\{{}$\1\6
${}\\{tom\_renew}(\\{tom},\39\\{mia}+\|i);{}$\6
\4${}\}{}$\2\6
\&{for} ${}(\|i\PP;{}$ ${}\|i<\.{MAX\_PROCS};{}$ ${}\|i\PP){}$\5
${}\{{}$\1\6
${}\\{tom\_renew}(\\{tom},\39\\{mia}+\|i);{}$\6
\4${}\}{}$\2\6
\4${}\}{}$\2\par
\U18.\fi

\M{21}A message modified the database, and that message was local,
i.e., generated on this node. To keep all instances of the
database up to date, we need to propagate this message to
all the other components:
\Y\B\4\X21:broadcast modifications to db\X${}\E{}$\6
${}\{{}$\1\6
\&{int} \|i;\7
\&{for} ${}(\|i\K\T{0};{}$ ${}\|i<\\{GetRoot}(\,);{}$ ${}\|i\PP){}$\5
${}\{{}$\1\6
${}\\{RemoteSendMessage}(\|i,\39\.{ALIAS}(\|i),\39{}$(\&{char} ${}{*}){}$ ${}{%
\AND}\\{message},\39{}$\&{sizeof} (\\{message}));\6
\4${}\}{}$\2\6
\&{for} ${}(\|i\PP;{}$ ${}\|i<\.{MAX\_PROCS};{}$ ${}\|i\PP){}$\5
${}\{{}$\1\6
${}\\{RemoteSendMessage}(\|i,\39\.{ALIAS}(\|i),\39{}$(\&{char} ${}{*}){}$ ${}{%
\AND}\\{message},\39{}$\&{sizeof} (\\{message}));\6
\4${}\}{}$\2\6
\4${}\}{}$\2\par
\Us18\ET37.\fi

\M{22}The just arrived message is of type \PB{\.{DB}}, i.e., it concerns
the database. Our copy of the database needs then to be updated.
\Y\B\4\X22:update your copy of the db\X${}\E{}$\6
\&{switch} ${}(\\{message}.\\{arg}[\T{0}]){}$\5
${}\{{}$\1\6
\4\&{case} \.{DB\_NEW\_STATUS}:\5
${}\\{db}.\\{node}[\\{message}.\\{subid}].\\{status}\K\\{message}.\\{arg}[%
\T{1}];{}$\6
\&{break};\6
\4\&{case} \.{DB\_NEW\_ROLE}:\5
${}\\{db}.\\{node}[\\{message}.\\{subid}].\\{role}\K\\{message}.\\{arg}[%
\T{1}];{}$\6
\&{break};\6
\4\&{case} \.{DB\_INC\_REBOOT}:\5
${}\\{db}.\\{node}[\\{message}.\\{subid}].\\{reboot\_nr}\PP;{}$\6
\&{break};\6
\4\&{case} \.{DB\_NEW\_TASK\_STATUS}:\5
${}\\{db}.\\{node}[\\{message}.\\{subid}].\\{task}[\\{message}.\\{arg}[\T{1}]].%
\\{status}\K\\{message}.\\{arg}[\T{2}];{}$\6
\&{break};\6
\4\&{case} \.{DB\_NEW\_TASK\_ERROR}:\5
${}\\{db}.\\{node}[\\{message}.\\{subid}].\\{task}[\\{message}.\\{arg}[\T{1}]].%
\\{status}\K\\{message}.\\{arg}[\T{2}];{}$\6
\&{break};\6
\4${}\}{}$\2\par
\Us18\ET37.\fi

\M{23}A \PB{\.{TAIA\_SEND\_TIMEOUT}} needs to be inserted again in the
timeout list.
\Y\B\4\X23:insert TAIA-timeout, subid=\PB{\\{subid}}\X${}\E{}$\6
$\\{tom\_insert}(\\{tom},\39\\{taia}+\\{message}.\\{subid}){}$;\par
\U18.\fi

\M{24}A node is up again. Broadcast the news
(of course, skipping yourself and the node back to life\dots)
\Y\B\4\X24:broadcast \PB{\.{NIUA}}\X${}\E{}$\6
${}\{{}$\1\6
\&{int} \|i${},{}$ \|n${},{}$ ${}\this{}$;\C{ note: message.subid is the id of
the Node which Is Up Again }\7
\&{for} ${}(\\{message}.\\{type}\K\.{NIUA},\39\|n\K\.{MAX\_PROCS},\39\|i\K%
\T{0},\39\this\K\\{GetRoot}(\,);{}$ ${}\|i<\|n;{}$ ${}\|i\PP){}$\1\6
\&{if} ${}(\|i\I\this\W\|i\I\\{message}.\\{subid}){}$\1\5
${}\\{RemoteSendMessage}(\|i,\39\.{ALIAS}(\|i),\39{}$(\&{char} ${}{*}){}$ ${}{%
\AND}\\{message},\39{}$\&{sizeof} (\\{message}));\2\2\6
\4${}\}{}$\2\par
\U18.\fi

\M{25}A node is up again, so I need to start again watching
its component.
\Y\B\4\X25:renew TAIA-timeout, \PB{\\{subid}}\X${}\E{}$\6
$\\{tom\_renew}(\\{tom},\39\\{taia}+\\{message}.\\{subid}){}$;\par
\U18.\fi

\M{26}We just entered a suspicion period---we need to discriminate
the case `{\it an agent is down\/}' from the case `{\it a complete
node is down\/}'. To do so, we start waiting for at most
\PB{\.{IMALIVE\_SET\_TIMEOUT}} for some signs of life coming from
the I'm Alive Task on node \PB{$\\{message}.\\{subid}$}. This is managed
bu simply inserting an $\underline{\hbox{acyclic}}$ timeout in the
timeout list as follows:
\Y\B\4\X26:insert \PB{\.{TEIF}} timeout\X${}\E{}$\6
$\\{tom\_declare}({\AND}\\{teif},\39\.{TEIF\_CYCLIC},\39\.{TOM\_SET\_ENABLE},%
\39\.{TEIF\_TIMEOUT},\39\\{message}.\\{subid},\39\.{IMALIVE\_SET\_TIMEOUT});{}$%
\6
${}\\{tom\_insert}(\\{tom},\39{\AND}\\{teif}){}$;\par
\Us18\ET37.\fi

\M{27}If we are suspecting an agent, there's no need to expect
something from it (dosn't it sound like philosophy? ;-)
so we need to suspend the cyclic \PB{\.{TAIA}} timeout; we do that
deleting it temporarily from the list.
\Y\B\4\X27:delete timeout (TAIA-timeout, \PB{\\{subid}})\X${}\E{}$\6
$\\{tom\_delete}(\\{tom},\39\\{taia}+\\{message}.\\{subid}){}$;\par
\U18.\fi

\M{28}Luckily, only the remote component is crashed, not the entire
node where it was running onto. This seems to be true because
a \PB{\.{TEIF}} message has been sent from the I'm Alive Task of
that node. The \PB{\.{TEIF}} timeout is consequently deleted.
\Y\B\4\X28:delete timeout (TEIF-timeout, \PB{\\{subid}})\X${}\E{}$\6
$\\{tom\_delete}(\\{tom},\39{\AND}\\{teif}){}$;\par
\U18.\fi

\M{29}Agent recovery will spawn a clone of the backup \PB{\\{subid}}.
If no backup clones are available, the entire node will
be rebooted. Now the problem is---who can do that? The
only component being alive on that node is\dots the I'm Alive
Task, so the only way to accomplish this task should be by
sending a ``spawn new component'' (\PB{\.{SPAN}}) message to that
I'm Alive Task:
\Y\B\4\X29:Agent-Recovery(\PB{\\{subid}})\X${}\E{}$\6
$\\{message}.\\{type}\K\.{SPAN};{}$\6
${}\\{RemoteSendMessage}(\\{message}.\\{subid},\39\.{IAT\_ALIAS},\39{}$(%
\&{char} ${}{*}){}$ ${}{\AND}\\{message},\39{}$\&{sizeof} (\\{message}));\par
\U18.\fi

\M{30}Same as above, but for the manager.
\Y\B\4\X30:Manager-Recovery(\PB{\\{managerid}})\X${}\E{}$\6
$\\{message}.\\{type}\K\.{SPAN};{}$\6
${}\\{RemoteSendMessage}(\\{managerid},\39\.{IAT\_ALIAS},\39{}$(\&{char}
${}{*}){}$ ${}{\AND}\\{message},\39{}$\&{sizeof} (\\{message}));\par
\U37.\fi

\M{31}This section covers the case no sign of life seems to
come from a suspected node. The node is therefore rebooted.
$\underline{\hbox{Open problem}}$: if TEX has a local scope, who
can do this? Probably this will require TEX to send a
message to its host via a (burst of) UDP writes; then
the host will take care of rebooting the node in question
or to do something else, e.g., triggering an alarm for
the operator. Anyway, for us this is simply a \PB{\\{TEXReboot}(\,)}.
\Y\B\4\X31:Node-Recovery(\PB{\\{subid}})\X${}\E{}$\C{
	TEXReboot(message.subid); 	   TEXReset(message.subid); 	}\par
\Us18\ET37.\fi

\M{32}Other messages are foreseen---their management {\bf will be}
put in here.
{\bf Note: for the moment, this section is shared amongst
manager and backups.}

\Y\B\4\X32:deal with these other messages\X${}\E{}$\6
${}\{{}$\1\6
\&{switch} ${}(\\{message}.\\{type}){}$\5
${}\{{}$\C{ ...to be added... }\6
\,${}\}{}$\C{ end switch }\6
\4${}\}{}$\2\par
\Us18\ET37.\fi

\M{33}A backup agent can be considered as a manager of a system
collapsing to its node. It only takes local-scope decisions
and actions. Nevertheless, quite a lot of its code is
inherited almost without modification from the Manager,
and its structure is basically the same of this latter.
\Y\B\4\X33:DIR net backup agent\X${}\E{}$\6
${}\{{}$\1\6
${}\\{LogError}(\.{EC\_ERROR},\39\.{"Backup"},\39\.{"Backup\ starts..."}){}$;%
\C{ the alarm of these timeouts simply sends a message of id 	
<timeout-type>, subid = \PB{\\{subid}} to the backup agent 	 }\6
\X34:insert four timeouts (IA-flag, MIA, TAIA, and TEIF)\X\C{
	TimeWaitHigh(TimeNowHigh() + 300000); 	}\C{ suspicion period is set to off }\6
\X15:clear \PB{\\{suspicion\_period}}\X\6
\X16:clear IA-flag\X\6
${}\\{LogError}(\.{EC\_ERROR},\39\.{"Backup"},\39\.{"activating\ IAT..."});{}$\6
${}\\{LogError}(\.{EC\_ERROR},\39\.{"Backup"},\39\.{"Backup\ activates\ IA}\)%
\.{T..."});{}$\6
\X17:activate IAT\X\6
${}\\{LogError}(\.{EC\_ERROR},\39\.{"Backup"},\39\.{"Backup\ loop\ starts.}\)%
\.{.."});{}$\6
\X37:backup loop (waiting for incoming messages)\X\6
\4${}\}{}$\C{ end backup }\2\par
\U6.\fi

\M{34}The backup initialises a set of timeout objects and inserts
them into the timeout list, more or less the way the manager does;
only, there's just one \PB{\.{MIA}} timeout comin' in and one \PB{\.{TAIA}}
coming out.
\Y\B\4\X34:insert four timeouts (IA-flag, MIA, TAIA, and TEIF)\X${}\E{}$\6
${}\{{}$\1\6
${}\\{tom}\K\\{tom\_init}(\\{send\_timeout\_message});{}$\6
\X35:declare and insert the one MIA timeout\X\6
\X12:declare and insert the IA-flag timeout\X\6
\X36:declare and insert the one TAIA timeout\X\6
\4${}\}{}$\2\par
\U33.\fi

\M{35}At most every \PB{\.{MIA\_DEADLINE\_B}} ticks a ``Manager Is Alive'' (%
\PB{\.{MIA}})
message needs to be received by a backup.
Note how, regardless the actual value of \PB{\\{managerid}}, the entry being
filled in is always entry 0.
\Y\B\4\X35:declare and insert the one MIA timeout\X${}\E{}$\6
${}\{{}$\1\6
${}\\{LogError}(\.{EC\_ERROR},\39\.{"backup"},\39\.{"managerid\ ==\ \%d"},\39%
\\{managerid});{}$\6
${}\\{tom\_declare}(\\{mia}+\\{managerid},\39\.{MIA\_CYCLIC\_B},\39\.{TOM\_SET%
\_ENABLE},\39\.{MIA\_TIMEOUT\_B},\39\\{managerid},\39\.{MIA\_DEADLINE\_B});{}$\6
${}\\{tom\_insert}(\\{tom},\39\\{mia}+\\{managerid});{}$\6
\\{tom\_dump}(\\{tom});\6
\4${}\}{}$\2\par
\U34.\fi

\M{36}At most every \PB{\.{TAIA\_DEADLINE\_B}} ticks a ``This Agent Is Alive''
(\PB{\.{TAIA}}) message needs to be sent from this backup to the
Manager.
Note how, regardless the actual value of \PB{\\{managerid}}, the entry being
filled in is always entry 0.
\Y\B\4\X36:declare and insert the one TAIA timeout\X${}\E{}$\6
${}\{{}$\1\6
${}\\{tom\_declare}(\\{taia},\39\.{TAIA\_CYCLIC\_B},\39\.{TOM\_SET\_ENABLE},\39%
\.{TAIA\_TIMEOUT\_B},\39\T{0},\39\.{TAIA\_DEADLINE\_B});{}$\6
${}\\{tom\_insert}(\\{tom},\39\\{taia});{}$\6
\4${}\}{}$\2\par
\U34.\fi

\M{37}This loop is the real core of the backup agent, the way
the manager loop was for the manager. It has do deal
with a number of messages coming from the timeout manager,
the manager, remote I'm Alive Tasks. As we said
in the corresponding section of the manager, this is
the core of the fault-tolerant strategy of the DIR net.
\Y\B\4\X37:backup loop (waiting for incoming messages)\X${}\E{}$\6
\&{while} (\T{1})\5
${}\{{}$\1\6
\X53:wait for an incoming message\X\6
${}\\{LogError}(\.{EC\_ERROR},\39\.{"Backup"},\39\.{"message\ received\ \%d}\)%
\.{\ [type\ ==\ \%s]\ from\ n}\)\.{ode\ \%d"},\39\\{message}.\\{type},\39%
\\{DIRPrintCode}(\\{message}.\\{type}),\39\\{message}.\\{local}\?\\{GetRoot}(%
\,):\\{message}.\\{subid});{}$\6
\&{switch} ${}(\\{message}.\\{type}){}$\5
${}\{{}$\1\6
\4\&{case} \.{IA\_FLAG\_TIMEOUT}:\C{ time to clear the IA-flag! }\6
\X16:clear IA-flag\X\6
${}\\{LogError}(\.{EC\_ERROR},\39\.{"Backup\ loop"},\39\.{"IA\_FLAG\_TIMEOUT\
->\ }\)\.{IA-flag\ cleared"});{}$\6
\&{break};\6
\4\&{case} \.{TAIA\_TIMEOUT\_B}:\C{ time to send a TAIA to the manager }\6
${}\\{LogError}(\.{EC\_ERROR},\39\.{"Backup\ loop"},\39\.{"TAIA\_TIMEOUT\_B\ ->%
\ s}\)\.{end\ TAIA\ to\ manager"});{}$\6
\X38:send \PB{\.{TAIA}} to the manager\X\6
\&{break};\6
\4\&{case} \.{DB}:\C{ a message which modifies the db }\6
${}\\{LogError}(\.{EC\_ERROR},\39\.{"Backup\ loop"},\39\.{"DB\ message."});{}$\6
\&{if} ${}(\\{message}.\\{subid}\E\\{GetRoot}(\,){}$)\C{ the message is local }%
\6
${}\{{}$\1\6
\X22:update your copy of the db\X\6
\X21:broadcast modifications to db\X\C{ this is sent also to the manager,
therefore 			   it is a TAIA in piggybacking. 			 }\6
\X39:renew \PB{\.{TAIA}} timeout\X\6
\4${}\}{}$\2\6
\&{else}\5
${}\{{}$\1\6
\X22:update your copy of the db\X\6
\&{if} ${}(\\{message}.\\{subid}\E\\{managerid}){}$\1\5
${}\\{tom\_renew}(\\{tom},\39\\{mia}+\\{managerid});{}$\2\6
\4${}\}{}$\2\6
\&{break};\6
\4\&{case} \.{MIA}:\5
${}\\{LogError}(\.{EC\_ERROR},\39\.{"Backup\ loop"},\39\.{"MIA\ message:\
manage}\)\.{rid==\%d,\ arg[0]==\%d,}\)\.{\ \%s."},\39\\{managerid},\39%
\\{message}.\\{arg}[\T{0}],\39(\\{managerid}\E\\{message}.\\{arg}[\T{0}])\?%
\.{"equal"}:\.{"different"}){}$;\C{ if a MIA comes in... }\6
\&{if} ${}(\R\\{tom\_ispresent}(\\{tom},\39\\{mia}+\\{message}.\\{arg}[%
\T{0}])){}$\5
${}\{{}$\1\6
${}\\{LogError}(\.{EC\_ERROR},\39\.{"Backup\ loop"},\39\.{"MIA\ timeout\ \%d\
is\ n}\)\.{ot\ present!"},\39\\{message}.\\{arg}[\T{0}]);{}$\6
\\{tom\_dump}(\\{tom});\C{ a new manager has been chosen }\6
${}\\{tom\_delete}(\\{tom},\39\\{mia}+\\{managerid});{}$\6
${}\\{LogError}(\.{EC\_ERROR},\39\.{"Backup\ loop"},\39\.{"MIA\ timeout\ \%d\
dele}\)\.{ted!"},\39\\{managerid});{}$\6
${}\\{managerid}\K\\{message}.\\{arg}[\T{0}];{}$\6
${}\\{LogError}(\.{EC\_ERROR},\39\.{"Backup\ loop"},\39\.{"New\ manager\ is\ %
\%d"},\39\\{managerid});{}$\6
\4${}\}{}$\2\6
${}\\{tom\_renew}(\\{tom},\39\\{mia}+\\{managerid}){}$;\C{ if you get a MIA
while expecting a TEIF, then 		   no problem, simply get out of the suspicion
period 		 }\6
\&{if} ${}({*}\\{suspicion\_period}\E\.{TRUE}){}$\5
${}\{{}$\1\6
${}{*}\\{suspicion\_period}\K\.{FALSE};{}$\6
${}\\{LogError}(\.{EC\_ERROR},\39\.{"Backup"},\39\.{"got\ a\ MIA\ while\ exp}\)%
\.{ecting\ a\ TEIF\ --\ got}\)\.{\ out\ of\ the\ suspicio}\)\.{n\
period!"});{}$\6
\4${}\}{}$\2\6
\&{break};\6
\4\&{case} \.{MIA\_TIMEOUT\_B}:\5
${}\\{LogError}(\.{EC\_ERROR},\39\.{"Backup\ loop"},\39\.{"MIA\_TIMEOUT\_B\
messa}\)\.{ge:\ no\ heartbeat\ fro}\)\.{m\ the\ manager\ --\ sus}\)\.{picion\
period\ entere}\)\.{d"}){}$;\C{ no heartbeat from the manager... enter 		   a
suspicion period then 		 }\6
${}{*}\\{suspicion\_period}\K\.{TRUE};{}$\6
${}\\{message}.\\{subid}\K\\{managerid};{}$\6
${}\\{LogError}(\.{EC\_ERROR},\39\.{"Backup\ loop"},\39\.{"About\ to\ insert\ a%
\ T}\)\.{EIF"});{}$\6
\X26:insert \PB{\.{TEIF}} timeout\X\6
${}\\{LogError}(\.{EC\_ERROR},\39\.{"Backup\ loop"},\39\.{"TEIF\
inserted"});{}$\6
${}\\{tom\_delete}(\\{tom},\39\\{mia}+\\{managerid});{}$\6
\&{break};\6
\4\&{case} \.{TEIF}:\5
${}\\{LogError}(\.{EC\_ERROR},\39\.{"Backup\ loop"},\39\.{"TEIF\ message\ --\
rec}\)\.{eived\ a\ message\ from}\)\.{\ IAT@node\%d"},\39\\{message}.%
\\{subid}){}$;\C{ a TEIF message has been sent from a IAT: 		   as its last
action, the IAT went to sleep 		 }\6
\&{if} ${}({*}\\{suspicion\_period}\E\.{TRUE}\W\\{message}.\\{subid}\E%
\\{managerid}){}$\5
${}\{{}$\C{ 			delete timeout (MIA-timeout, \PB{\\{subid}}); 			tom%
\_delete(tom, mia + managerid); 			}\1\6
${}\\{tom\_delete}(\\{tom},\39{\AND}\\{teif});{}$\6
${}{*}\\{suspicion\_period}\K\.{FALSE};{}$\6
${}\\{LogError}(\.{EC\_ERROR},\39\.{"Backup\ loop"},\39\.{"Manager\ needs\ to\
be}\)\.{\ recovered!"}){}$;\C{ Manager recovery will spawn a clone of the 			
backup \PB{\\{subid}}. If no more manager clones 			   are available, the
entire node will 			   be rebooted. 			 }\6
\X30:Manager-Recovery(\PB{\\{managerid}})\X\6
\4${}\}{}$\2\6
\&{else}\5
${}\{{}$\1\6
${}\\{LogError}(\.{EC\_ERROR},\39\.{"Backup\ loop"},\39\.{"IAT@node\%d\ needs\
to}\)\.{\ be\ awaken\ --\ sent\ E}\)\.{NIA\ to\ component@nod}\)\.{e\%d"},\39%
\\{message}.\\{subid},\39\\{message}.\\{subid}){}$;\C{ send ENIA (i.e.,
``ENable IAt'') to \PB{\\{subid}} }\6
${}\\{message}.\\{type}\K\.{ENIA};{}$\6
${}\\{RemoteSendMessage}(\\{message}.\\{subid},\39\.{ALIAS}(\\{message}.%
\\{subid}),\39{}$(\&{char} ${}{*}){}$ ${}{\AND}\\{message},\39{}$\&{sizeof} (%
\\{message}));\6
${}\\{tom\_renew}(\\{tom},\39\\{mia}+\\{managerid});{}$\6
\4${}\}{}$\2\6
\&{break};\6
\4\&{case} \.{TEIF\_TIMEOUT\_B}:\5
${}\\{LogError}(\.{EC\_ERROR},\39\.{"Backup\ loop"},\39\.{"TEIF\_TIMEOUT\_B\
mess}\)\.{age."});{}$\6
\&{if} ${}({*}\\{suspicion\_period}\E\.{TRUE}){}$\5
${}\{{}$\1\6
${}\\{LogError}(\.{EC\_ERROR},\39\.{"Backup\ loop"},\39\.{"the\ node\ is\
suspect}\)\.{ed\ =>\ has\ crashed."});{}$\6
\X40:delete MIA-timeout coming from \PB{\\{subid}}, if any exists\X\6
${}{*}\\{suspicion\_period}\K\.{FALSE}{}$;\C{ the entire node will be rebooted.
			 }\6
${}\\{LogError}(\.{EC\_ERROR},\39\.{"Backup\ loop"},\39\.{"node\
recovery!"});{}$\6
\X31:Node-Recovery(\PB{\\{subid}})\X\6
${}\\{LogError}(\.{EC\_ERROR},\39\.{"Backup\ loop"},\39\.{"`a\ node\ is\ down'\
me}\)\.{ssage\ sent\ around"});{}$\6
\X41:send \PB{\.{ANID}} to all except \PB{\\{managerid}}\X\6
${}\\{LogError}(\.{EC\_ERROR},\39\.{"Backup\ loop"},\39\.{"choice\ of\ new\
manag}\)\.{er"});{}$\6
\X42:choose next manager\X\C{ if this backup is to be the new manager... }\6
\&{if} ${}(\\{managerid}\E\\{GetRoot}(\,)){}$\5
${}\{{}$\1\6
${}\\{mystate}.\\{role}\K\.{DIR\_MANAGER};{}$\6
${}\\{TEXSetState}({\AND}\\{mystate});{}$\6
\\{TEXRestartTask}(\,);\6
\4${}\}{}$\2\6
\4${}\}{}$\2\6
\&{else}\5
${}\{{}$\1\6
${}\\{LogError}(\.{EC\_ERROR},\39\.{"Backup\ loop"},\39\.{"IAT\ of\ the\
manager\ }\)\.{needs\ to\ be\ awaken\ -}\)\.{-\ sent\ ENIA\ to\ Manag}\)%
\.{er@node\%d"},\39\\{message}.\\{subid}){}$;\C{ send ENIA (i.e., ``ENable
IAt'') to <managerid> }\6
${}\\{message}.\\{type}\K\.{ENIA};{}$\6
${}\\{RemoteSendMessage}(\\{managerid},\39\.{ALIAS}(\\{managerid}),\39{}$(%
\&{char} ${}{*}){}$ ${}{\AND}\\{message},\39{}$\&{sizeof} (\\{message}));\C{
renew MIA-timeout }\6
${}\\{tom\_renew}(\\{tom},\39\\{mia}+\\{managerid});{}$\6
\4${}\}{}$\2\6
\&{break};\6
\4\&{case} \.{WITM}:\6
${}\\{LogError}(\.{EC\_ERROR},\39\.{"Backup\ loop"},\39\.{"WITM\
message."});{}$\6
${}\\{sender}\K\\{message}.\\{subid};{}$\6
${}\\{message}.\\{arg}[\T{0}]\K\\{managerid};{}$\6
${}\\{message}.\\{subid}\K\\{GetRoot}(\,);{}$\6
${}\\{message}.\\{type}\K\.{NMI};{}$\6
${}\\{RemoteSendMessage}(\\{sender},\39\.{ALIAS}(\\{sender}),\39{}$(\&{char}
${}{*}){}$ ${}{\AND}\\{message},\39{}$\&{sizeof} (\\{message}));\6
\&{break};\6
\4\&{case} \.{ENIA}:\5
${}\\{LogError}(\.{EC\_ERROR},\39\.{"Backup\ loop"},\39\.{"ENIA\
message."});{}$\6
\X16:clear IA-flag\X\C{ if an IAT gets an ``activate'' message while it's 		
active, the message is ignored }\6
\X17:activate IAT\X\6
\&{break};\6
\4\&{case} \.{NMI}:\5
${}\\{LogError}(\.{EC\_ERROR},\39\.{"Backup\ loop"},\39\.{"NMI\ message."});{}$%
\6
\&{if} ${}(\\{message}.\\{arg}[\T{0}]\E\\{managerid}){}$\1\5
\&{break};\C{ something is going wrong -- someone has 		   a different
managerid }\2\6
${}\\{LogError}(\.{EC\_ERROR},\39\.{"Backup\ loop"},\39\.{"node\ \%d\ thinks\
the\ }\)\.{manager\ is\ \%d,\ while}\)\.{\ I\ think\ it\ is\ \%d."},\39%
\\{message}.\\{subid},\39\\{message}.\\{arg}[\T{0}],\39\\{managerid});{}$\6
\&{break};\6
\4\&{case} \.{REQUEST\_DB}:\6
${}\\{LogError}(\.{EC\_ERROR},\39\.{"Backup\ loop"},\39\.{"REQUEST\_DB\
message.}\)\.{"}){}$;\C{ a node is requesting a full copy of the database }\6
${}\\{RemoteSendMessage}(\\{message}.\\{subid},\39\.{ALIAS}(\\{message}.%
\\{subid}),\39{}$(\&{char} ${}{*}){}$ ${}{\AND}\\{db},\39{}$\&{sizeof} (%
\\{db}));\6
\&{break};\6
\4\&{default}:\5
${}\\{LogError}(\.{EC\_ERROR},\39\.{"Backup\ loop"},\39\.{"other\
messages."});{}$\6
\X32:deal with these other messages\X\6
\4${}\}{}$\C{ end switch (message.type) }\2\6
\X16:clear IA-flag\X\6
\X43:renew IA-flag-timeout\X\6
\4${}\}{}$\C{ end backup loop }\2\par
\U33.\fi

\M{38}A ``This Agent Is Alive'' message needs to be sent to the manager.
\Y\B\4\X38:send \PB{\.{TAIA}} to the manager\X${}\E{}$\6
$\\{message}.\\{type}\K\.{TAIA};{}$\6
${}\\{message}.\\{subid}\K\\{GetRoot}(\,);{}$\6
${}\\{LogError}(\.{EC\_ERROR},\39\.{"Backup"},\39\.{"sending\ TAIA\ to\ man}\)%
\.{ager..."});{}$\6
${}\\{RemoteSendMessage}(\\{managerid},\39\.{ALIAS}(\\{managerid}),\39{}$(%
\&{char} ${}{*}){}$ ${}{\AND}\\{message},\39{}$\&{sizeof} (\\{message}));\6
${}\\{LogError}(\.{EC\_ERROR},\39\.{"Backup"},\39\.{"...TAIA\ sent\ to\ man}\)%
\.{ager."}){}$;\par
\U37.\fi

\M{39}The local database has been modified---this calls for
a broadcast of the modifications. One of those who will
receive these modifications is the manager.
Such information holds implicitly (in piggybacking) a TAIA,
so we can renew that timeout.
\Y\B\4\X39:renew \PB{\.{TAIA}} timeout\X${}\E{}$\6
$\\{tom\_renew}(\\{tom},\39\\{taia}){}$;\par
\U37.\fi

\M{40}A \PB{\.{TEIF\_TIMEOUT\_B}} has come from node \PB{$\\{message}.%
\\{subid}$}. Here
we build a timeout object and try to delete an entry pertaining
node \PB{$\\{message}.\\{subid}$} and holding a \PB{\.{MIA\_TIMEOUT\_B}}.
\Y\B\4\X40:delete MIA-timeout coming from \PB{\\{subid}}, if any exists\X${}%
\E{}$\6
${}\{{}$\1\6
\&{timeout\_t} \|t;\7
${}\\{tom\_declare}({\AND}\|t,\39\.{MIA\_CYCLIC\_B},\39\.{TOM\_SET\_ENABLE},\39%
\.{MIA\_TIMEOUT\_B},\39\\{message}.\\{subid},\39\.{MIA\_DEADLINE\_B});{}$\6
${}\\{tom\_delete}(\\{tom},\39{\AND}\|t);{}$\6
\4${}\}{}$\2\par
\U37.\fi

\M{41}``A Node Is Down'' (\PB{\.{ANID}}) is sent to everyone but the down node.
\Y\B\4\X41:send \PB{\.{ANID}} to all except \PB{\\{managerid}}\X${}\E{}$\6
${}\{{}$\1\6
\&{int} \|i${},{}$ \|n${},{}$ ${}\this;{}$\7
\&{for} ${}(\|i\K\T{0},\39\\{message}.\\{type}\K\.{ANID},\39\|n\K\.{MAX%
\_PROCS},\39\this\K\\{GetRoot}(\,);{}$ ${}\|i<\|n;{}$ ${}\|i\PP){}$\1\6
\&{if} ${}(\|i\I\\{managerid}\W\|i\I\this){}$\1\5
${}\\{RemoteSendMessage}(\|i,\39\.{ALIAS}(\|i),\39{}$(\&{char} ${}{*}){}$ ${}{%
\AND}\\{message},\39{}$\&{sizeof} (\\{message}));\2\2\6
\4${}\}{}$\2\par
\U37.\fi

\M{42}This is the na\"ivest strategy for choosing a new manager.
Some more sophisticated (and safer) protocol will be used in
the future.
\Y\B\4\X42:choose next manager\X${}\E{}$\6
$\\{managerid}\PP;{}$\6
\&{if} ${}(\\{managerid}\G\.{MAX\_PROCS}){}$\1\5
${}\\{managerid}\K\T{0}{}$;\2\par
\U37.\fi

\M{43}Renew the IA-flag timeout.
\Y\B\4\X43:renew IA-flag-timeout\X${}\E{}$\6
$\\{tom\_renew}(\\{tom},\39{\AND}\\{ia}){}$;\par
\Us18\ET37.\fi

\M{44}The number of tasks on this node is stored in the database.
\Y\B\4\X44:store number of tasks (\PB{\|n});\X${}\E{}$\6
${}\{{}$\1\6
${}\\{db}.\\{node}[\\{GetRoot}(\,)].\\{task\_nr}\K\|n;{}$\6
\4${}\}{}$\2\par
\U8.\fi

\M{45}Information pertaining task \PB{\|i} is stored in the database.
\Y\B\4\X45:store in local database(i, status)\X${}\E{}$\6
${}\{{}$\1\6
\&{int} \\{this\_node}${}\K\\{GetRoot}(\,){}$;\C{ is thread i running or
waiting? Is it isolated / faulty / ok... }\7
${}\\{db}.\\{node}[\\{this\_node}].\\{task}[\|i].\\{status}\K\\{status}{}$;\C{
initialize error\_nb to zero }\6
${}\\{db}.\\{node}[\\{this\_node}].\\{task}[\|i].\\{error\_nr}\K\T{0};{}$\6
\4${}\}{}$\2\par
\U8.\fi

\M{46}This section runs the algorithm used in the Voting Farm
to manage the problem of global broadcasts (each component
has to broadcast some data to all the others and has to
deliver data sent by all the others). See also ``the Algorithm
of Pipelined Broadcast''.
\Y\B\4\X46:broadcast local database and receive the others' databases\X${}\E{}$%
\6
${}\{{}$\1\6
\&{int} \\{nProcs}${}\K\.{MAX\_PROCS};{}$\6
\&{int} \\{task\_nr}${},{}$ \|i;\7
\&{for} ${}(\|i\K\T{0};{}$ ${}\|i<\\{nProcs};{}$ ${}\|i\PP){}$\5
${}\{{}$\1\6
\&{if} ${}(\|i\E\\{GetRoot}(\,)){}$\5
${}\{{}$\1\6
\X48:broadcast the local part of the database\X\6
\4${}\}{}$\2\6
\&{else}\5
${}\{{}$\1\6
\X47:deliver remote database\X\6
\4${}\}{}$\2\6
\4${}\}{}$\2\6
\4${}\}{}$\2\par
\U8.\fi

\M{47}A remote database is delivered in three steps: first, the sender sends
an integer with its node-id; second, the number of tasks to be transfered
is sent; and third, the task information is received into the proper `slot'.
The first two integers are received in the node's ``normal'' mailbox,
the bulk of data is received via the \PB{\.{DB\_MBOX}} mailbox.
Errors are also sent, if any.
\Y\B\4\X47:deliver remote database\X${}\E{}$\6
$\{{}$\6
\&{int} \|n;\7
${}\|n\K\&{sizeof}(\&{int});{}$\6
${}\\{TEXReceiveMessage}(\.{MBOX}(\\{GetRoot}(\,)),\39{}$(\&{char} ${}{*}){}$
${}{\AND}\\{sender},\39{\AND}\|n,\39\.{INFINITE});{}$\6
\\{fflush}(\\{stdout});\6
${}\|n\K\&{sizeof}(\&{int});{}$\6
${}\\{TEXReceiveMessage}(\.{MBOX}(\\{GetRoot}(\,)),\39{}$(\&{char} ${}{*}){}$
${}{\AND}\\{task\_nr},\39{\AND}\|n,\39\.{INFINITE});{}$\6
\&{if} (\\{task\_nr})\5
${}\{{}$\1\6
${}\|n\K\\{task\_nr}*{}$\&{sizeof} (\\{DIR\_task\_t});\6
${}\\{TEXReceiveMessage}(\.{DB\_MBOX},\39{}$(\&{char} ${}{*}){}$ \\{db}${}.%
\\{node}[\\{sender}].\\{task},\39{\AND}\|n,\39\.{INFINITE});{}$\6
\4${}\}{}$\2\6
${}\\{db}.\\{node}[\\{sender}].\\{task\_nr}\K\\{task\_nr}{}$;\6
${}\|n\K\&{sizeof}(\&{int});{}$\6
${}\\{TEXReceiveMessage}(\.{MBOX}(\\{GetRoot}(\,)),\39{}$(\&{char} ${}{*}){}$
${}{\AND}\\{db}.\\{node}[\\{sender}].\\{error\_nr},\39{\AND}\|n,\39%
\.{INFINITE});{}$\6
\&{if} ${}(\\{db}.\\{node}[\\{sender}].\\{error\_nr})$ $\{{}$\6
$\|n\K\\{db}.\\{node}[\\{sender}].\\{error\_nr}*{}$\&{sizeof} (\\{DIR\_task%
\_t});\6
\\{TEXReceiveMessage} $(\.{DB\_MBOX},\39$ (\&{char} ${}{*}){}$ \\{db}${}.%
\\{node}[\\{sender}]$ $.$ \&{error} $,$ ${\AND}\|n,\39\.{INFINITE}$ )  ;\6
$\}{}$\6
$\}{}$\par
\U46.\fi

\M{48}This is symmetrical to the previous section. The order of broadcast is
optimal with respect to maximizing the throughput of a fully connected
(crossbar)
system (see ``{\it the Algorithm of Pipelined Broadcast\/}'').
The first two integers are sent to the destination node's ``normal'' mailbox,
the bulk of data is sent to the destination node's \PB{\.{DB\_MBOX}} mailbox.
Error informationation is also sent the same way.
\Y\B\4\X48:broadcast the local part of the database\X${}\E{}$\6
$\{$ \&{int} \|i${},{}$ \\{nprocs}${},{}$ \\{me};\7
${}\\{me}\K\\{GetRoot}(\,);{}$\6
${}\\{nprocs}\K\.{MAX\_PROCS};$ \&{for} ${}(\|i\K\T{0};{}$ ${}\|i<%
\\{nprocs};{}$ ${}\|i\PP)$ $\{$ \6
\&{if} ${}(\|i\E\\{me}){}$\1\5
\&{continue};\2\6
${}\\{RemoteSendMessage}(\|i,\39\.{ALIAS}(\|i),\39{}$(\&{char} ${}{*}){}$ ${}{%
\AND}\\{me},\39\&{sizeof}(\&{int}));{}$\6
${}\\{RemoteSendMessage}(\|i,\39\.{ALIAS}(\|i),\39{}$(\&{char} ${}{*}){}$ ${}{%
\AND}\\{db}.\\{node}[\\{me}].\\{task\_nr},\39\&{sizeof}(\&{int}));{}$\6
\&{if} ${}(\\{db}.\\{node}[\\{me}].\\{task\_nr}){}$\5
${}\{{}$\1\6
${}\\{RemoteSendMessage}(\|i,\39\.{DB\_MBOX},\39{}$(\&{char} ${}{*}){}$ %
\\{db}${}.\\{node}[\\{me}].\\{task},\39\\{db}.\\{node}[\\{me}].\\{task\_nr}*{}$%
\&{sizeof} (\\{DIR\_task\_t}));\6
\4${}\}{}$\2\6
${}\\{RemoteSendMessage}(\|i,\39\.{ALIAS}(\|i),\39{}$(\&{char} ${}{*}){}$ ${}{%
\AND}\\{db}.\\{node}[\\{me}].\\{error\_nr},\39\&{sizeof}(\&{int}));$ \&{if}
${}(\\{db}.\\{node}[\\{me}].\\{error\_nr})$ $\{$ \\{RemoteSendMessage} $(\|i,%
\39\.{DB\_MBOX},\39$ (\&{char} ${}{*}){}$ \\{db}${}.\\{node}[\\{me}]$ $.$ %
\&{error} $,$ $(\\{db}.\\{node}[\\{me}].\\{error\_nr})*{}$(\&{sizeof} (\\{DIR%
\_error\_t})) )  ; $\}$ $\}$ $\}{}$\par
\U46.\fi

\M{49}Here we fill in with default values the ``dynamic'' (modifiable) part
of the database, i.e., errors and the like.
\Y\B\4\X49:build a global database\X${}\E{}$\6
${}\{{}$\1\6
\&{int} \\{nProcs}${}\K\.{MAX\_PROCS};{}$\7
\&{for} ${}(\|i\K\T{0};{}$ ${}\|i<\\{nProcs};{}$ ${}\|i\PP){}$\5
${}\{{}$\1\6
${}\\{db}.\\{node}[\|i].\\{error\_nr}\K\T{0};{}$\6
${}\\{db}.\\{node}[\|i].\\{update\_nr}\K\T{0};{}$\6
${}\\{db}.\\{node}[\|i].\\{reboot\_nr}\K\T{0};{}$\6
\4${}\}{}$\2\6
\4${}\}{}$\2\par
\U8.\fi

\M{50}The whole database needs to survive to reboots. This is accomplished
by means of the \PB{\\{DataReset}(\,)} function, which marks data as
`reboot-resistant'.
\Y\B\4\X50:mark as `reboot-resistant' the whole database via \PB{\\{DataReset}(%
\,)}\X${}\E{}$\6
$\{{}$\C{ TXT has to supply this information }\6
\,${}\}{}$\par
\U8.\fi

\M{51}Sends a \PB{\.{REQUEST\_DB}} message to one or more of the
neighbouring dirnet components, and waits until it gets a full copy
of the global db. Note: it uses a special mailbox for that,
because the size of those messages is extremely larger with respect
to the size of messages for `normal' mailboxes.
Each node, say node $n$, sends requests first to node $n+1 \,\hbox{mod}\, \PB{%
\.{MAX\_PROCS}}$,
on a circular basis.
\Y\B\4\X51:request a copy of the global database\X${}\E{}$\6
${}\{{}$\1\6
\&{int} \\{this\_node}${}\K\\{GetRoot}(\,);{}$\6
\&{int} \\{nProcs}${}\K\.{MAX\_PROCS};{}$\6
\&{int} \|i;\6
\&{int} \\{size};\6
\&{int} \\{retval};\7
${}\\{message}.\\{type}\K\.{REQUEST\_DB};{}$\6
${}\\{message}.\\{subid}\K\\{this\_node};{}$\6
\&{for} ${}(\|i\K\\{this\_node}+\T{1};{}$ ${}\|i<\\{nProcs};{}$ ${}\|i\PP){}$\5
${}\{{}$\1\6
${}\\{RemoteSendMessage}(\|i,\39\.{ALIAS}(\|i),\39{}$(\&{char} ${}{*}){}$ ${}{%
\AND}\\{message},\39{}$\&{sizeof} (\\{message}));\6
${}\\{size}\K{}$\&{sizeof} (\\{db});\6
${}\\{retval}\K\\{TEXReceiveMessage}(\.{DB\_MBOX},\39{}$(\&{char} ${}{*}){}$
${}{\AND}\\{db},\39{\AND}\\{size},\39\.{REPLY\_DB\_TIMEOUT});{}$\6
\&{if} ${}(\\{retval}\E\.{MSG\_OK}){}$\1\5
\&{break};\2\6
\4${}\}{}$\2\6
\&{if} ${}(\\{retval}\I\.{MSG\_OK}){}$\1\6
\&{for} ${}(\|i\K\T{0};{}$ ${}\|i<\\{this\_node};{}$ ${}\|i\PP){}$\5
${}\{{}$\1\6
${}\\{RemoteSendMessage}(\|i,\39\.{ALIAS}(\|i),\39{}$(\&{char} ${}{*}){}$ ${}{%
\AND}\\{message},\39{}$\&{sizeof} (\\{message}));\6
${}\\{size}\K{}$\&{sizeof} (\\{db});\6
${}\\{retval}\K\\{TEXReceiveMessage}(\.{DB\_MBOX},\39{}$(\&{char} ${}{*}){}$
${}{\AND}\\{db},\39{\AND}\\{size},\39\.{REPLY\_DB\_TIMEOUT});{}$\6
\4${}\}{}$\2\2\6
\4${}\}{}$\2\par
\U8.\fi

\M{52}This is the function which is called when a timeout
occurs. In a sense, it translates a timeout event into
a message event which is sent to the local DIR net component.
\Y\B\4\X52:Alarm function\X${}\E{}$\6
\&{int} \\{send\_timeout\_message}(\&{TOM} ${}{*}\\{tom}){}$\1\1\2\2\6
${}\{{}$\1\6
${}\\{message}.\\{type}\K\\{tom}\MG\\{top}\MG\\{timeout}.\\{id};{}$\6
${}\\{message}.\\{subid}\K\\{tom}\MG\\{top}\MG\\{timeout}.\\{subid};{}$\6
${}\\{message}.\\{local}\K\T{1};{}$\6
\&{return} \\{TEXSendMessage}${}(\.{MBOX}(\\{GetRoot}(\,)),\39{}$(\&{char}
${}{*}){}$ ${}{\AND}\\{message},\39{}$\&{sizeof} (\\{message}));\C{ no test on
return value at the moment }\6
\4${}\}{}$\2\par
\U4.\fi

\M{53}Note: at the moment, this is shared among backup and
manager! The message should be received into a global
variable called ``message'', holding fields like
``type'', ``subid'', ``arg[0..argc]''\dots
Open question: {\it is it possible to specify a loop like the following one,
or is it better to have even a small timeout?}
\Y\B\4\X53:wait for an incoming message\X${}\E{}$\6
${}\{{}$\1\6
\&{int} \\{size}${}\K\&{sizeof}(\&{message\_t});{}$\7
${}\\{LogError}(\.{EC\_MESS},\39\.{"<wait\ for\ an\ incomi}\)\.{ng\ message>"},%
\39\.{"waiting..."});{}$\6
${}\\{TEXReceiveMessage}(\.{MBOX}(\\{GetRoot}(\,)),\39{}$(\&{char} ${}{*}){}$
${}{\AND}\\{message},\39{\AND}\\{size},\39\.{INFINITE});{}$\6
\4${}\}{}$\2\par
\Us18\ET37.\fi

\M{54}Broadcast a `Who Is The Manager' (\PB{\.{WITM}}) message.
\Y\B\4\X54:send \PB{\.{WITM}} to all\X${}\E{}$\6
${}\{{}$\1\6
\&{int} \|i;\7
${}\\{message}.\\{type}\K\.{WITM};{}$\6
\&{for} ${}(\|i\K\T{0};{}$ ${}\|i<\\{GetRoot}(\,);{}$ ${}\|i\PP){}$\5
${}\{{}$\1\6
${}\\{RemoteSendMessage}(\|i,\39\.{ALIAS}(\|i),\39{}$(\&{char} ${}{*}){}$ ${}{%
\AND}\\{message},\39{}$\&{sizeof} (\\{message}));\6
\4${}\}{}$\2\6
\&{for} ${}(\|i\PP;{}$ ${}\|i<\.{MAX\_PROCS};{}$ ${}\|i\PP){}$\5
${}\{{}$\1\6
${}\\{RemoteSendMessage}(\|i,\39\.{ALIAS}(\|i),\39{}$(\&{char} ${}{*}){}$ ${}{%
\AND}\\{message},\39{}$\&{sizeof} (\\{message}));\6
\4${}\}{}$\2\6
\4${}\}{}$\2\par
\U8.\fi

\M{55}If a node has rebooted or a new `generic component' took
the place of another one, then the running component needs
to clarify its own role in the dirnet. This is done by
sending a \PB{\.{WITM}} around. The following step waits until
some \PB{\.{NMI}} (`New Manager Is\dots') messages come in.
\Y\B\4\X55:wait for \PB{\.{NMI}} messages to come\X${}\E{}$\6
${}\{{}$\1\6
\&{int} \|i${},{}$ \|n${},{}$ \\{size};\7
\&{for} ${}(\|i\K\T{0},\39\|n\K\.{MAX\_PROCS},\39\\{size}\K\&{sizeof}(%
\&{message\_t});{}$ \T{1};\C{ forever }\6
${}\|i\K\PP\|i\MOD\|n){}$\1\6
\&{if} ${}(\\{TEXReceiveMessage}(\.{MBOX}(\|i),\39{}$(\&{char} ${}{*}){}$ ${}{%
\AND}\\{message},\39{\AND}\\{size},\39\T{0})\E\.{MSG\_OK}){}$\1\6
\&{if} ${}(\\{message}.\\{type}\E\.{NMI}){}$\1\5
\&{break};\2\6
\&{else}\5
${}\{{}$\1\6
${}\\{LogError}(\.{EC\_ERROR},\39\.{"Backup"},\39\.{"NMI\ message\ expecte}\)%
\.{d,\ \%d\ message\ receiv}\)\.{ed\ from\ node\ \%d"},\39\\{message}.\\{type},%
\39\|i);{}$\6
\4${}\}{}$\2\2\2\6
\4${}\}{}$\2\par
\U8.\fi

\M{56}These two functions resp. get and set the overall state of this
generic component.
\Y\B\4\X56:\PB{\\{GetState}} and \PB{\\{SetState}}\X${}\E{}$\6
\\{TEXGetState}(\&{status\_t} ${}{*}\|s){}$\1\1\2\2\6
${}\{{}$\1\6
${}\\{memcpy}(\|s,\39{\AND}\\{db}.\\{status},\39\&{sizeof}(\&{status\_t})){}$;%
\C{ a dirty trick for the time being... }\6
${}\|s\MG\\{primary}\K\T{1};{}$\6
\4${}\}{}$\2\7
\\{TEXSetState}(\&{status\_t} ${}{*}\|s){}$\1\1\2\2\6
${}\{{}$\1\6
${}\\{memcpy}({\AND}\\{db}.\\{status},\39\|s,\39\&{sizeof}(\&{status\_t}));{}$\6
\4${}\}{}$\2\par
\U4.\fi

\M{57}If we are at true initialisation time, then the roles are read
from the RL script.

\Y\B\4\X57:Read your role from the RL script\X${}\E{}$\6
${}\{{}$\1\6
\&{int} \|i;\7
\&{for} ${}(\|i\K\T{0};{}$ ${}\|i<\.{RCODE\_CARD};{}$ ${}\|i\PP){}$\5
${}\{{}$\1\6
\&{if} ${}(\\{rcodes}[\|i][\T{0}]\E\.{R\_SET\_ROLE}\W\\{rcodes}[\|i][\T{1}]\E%
\\{GetRoot}(\,)){}$\5
${}\{{}$\1\6
\&{switch} (\\{rcodes}[\|i][\T{2}])\5
${}\{{}$\1\6
\4\&{case} \.{R\_AGENT}:\5
${}\\{db}.\\{role}\K\.{DIR\_AGENT};{}$\6
\&{break};\6
\4\&{case} \.{R\_BACKUP}:\5
${}\\{db}.\\{role}\K\.{DIR\_BACKUP};{}$\6
\&{break};\6
\4\&{case} \.{R\_MANAGER}:\5
${}\\{db}.\\{role}\K\.{DIR\_MANAGER};{}$\6
\&{break};\6
\4${}\}{}$\2\6
\&{break};\6
\4${}\}{}$\2\6
\4${}\}{}$\2\6
\4${}\}{}$\2\par
\U8.\fi

\M{58}If we are at true initialisation time, then the identity of
the manager is read from the RL script.

\Y\B\4\X58:Check who is the manager according to the RL script\X${}\E{}$\6
${}\{{}$\1\6
\&{int} \|i;\7
\&{for} ${}(\|i\K\T{0};{}$ ${}\|i<\.{RCODE\_CARD};{}$ ${}\|i\PP){}$\5
${}\{{}$\1\6
\&{if} ${}(\\{rcodes}[\|i][\T{0}]\E\.{R\_SET\_ROLE}\W\\{rcodes}[\|i][\T{2}]\E%
\.{R\_MANAGER}){}$\5
${}\{{}$\1\6
${}\\{managerid}\K\\{rcodes}[\|i][\T{1}];{}$\6
\&{break};\6
\4${}\}{}$\2\6
\4${}\}{}$\2\6
${}\\{LogError}(\.{EC\_ERROR},\39\.{"init"},\39\.{"the\ manager\ is\ on\ n}\)%
\.{ode\ \%d."},\39\\{managerid});{}$\6
\4${}\}{}$\2\par
\U8.\fi

\M{59}
\Y\B\4\X59:TEX routines simulated on EPX\X${}\E{}$\6
\8\#\&{ifndef} \.{TEX}\6
\\{RemoteSendMessage}(\&{int} \\{node}${},\39{}$\&{int} \\{alias}${},\39{}$%
\&{char} ${}{*}\\{message},\39{}$\&{int} \\{size})\1\1\2\2\6
${}\{{}$\1\6
\&{int} \\{retval};\6
\8\#\&{ifdef} \.{EPX\_MAILBOXES}\7
${}\\{LogError}(\.{EC\_DEBUG},\39\.{"sender"},\39\.{"about\ to\ send\ a\ \%d\ }%
\)\.{byte\ msg\ to\ node\ \%d,}\)\.{\ req-id\ \%d"},\39\\{size},\39\\{node},\39%
\\{alias});{}$\6
${}\\{retval}\K\\{PutMessage}(\\{node},\39\\{alias},\39\.{MSG\_TYPE\_USER%
\_START},\39\T{1},\39{-}\T{1},\39\\{message},\39\\{size});{}$\6
\&{if} ${}(\\{retval}<\T{0}){}$\5
${}\{{}$\1\6
\&{char} \\{st}[\T{80}];\7
${}\\{LogError}(\.{EC\_ERROR},\39\.{"RemoteSendMessage"},\39\.{"PutMessage\
error:"});{}$\6
${}\\{sprintf}(\\{st},\39\.{"PutMessage\ on\ node\ }\)\.{\%d:\ "},\39%
\\{GetRoot}(\,));{}$\6
\\{perror}(\\{st});\6
\4${}\}{}$\2\6
\8\#\&{else}\6
${}\\{retval}\K\\{SendNode}(\\{node},\39\\{alias},\39\\{message},\39%
\\{size});{}$\6
\&{if} ${}(\\{retval}<\T{0}){}$\5
${}\{{}$\1\6
${}\\{printe}(\.{"\%d:\ RemoteSendMessa}\)\.{ge:\\n"},\39\\{GetRoot}(\,));{}$\6
\\{perror}(\.{"Error\ sending\ with\ }\)\.{SendNode"});\6
\4${}\}{}$\2\6
\&{else} \&{if} ${}(\\{retval}>\T{0}){}$\5
${}\{{}$\1\6
${}\\{printe}(\.{"\%d:\ RemoteSendMessa}\)\.{ge:\\n"},\39\\{GetRoot}(\,));{}$\6
${}\\{printe}(\.{"\%d:\ send\ message\ si}\)\.{ze\ \ is\ \ greater\ \ tha}\)%
\.{n\ \ the\ receive\ buffe}\)\.{r\\n"},\39\\{GetRoot}(\,));{}$\6
\4${}\}{}$\2\6
\8\#\&{endif}\6
\4${}\}{}$\2\7
\\{TEXSendMessage}(\&{int} \\{mbox}${},\39{}$\&{char} ${}{*}\\{message},\39{}$%
\&{int} \\{size})\1\1\2\2\6
${}\{{}$\1\6
\&{int} \\{retval};\6
\8\#\&{ifdef} \.{EPX\_MAILBOXES}\7
${}\\{LogError}(\.{EC\_DEBUG},\39\.{"sender"},\39\.{"about\ to\ send\ a\ \%d\ }%
\)\.{byte\ msg\ to\ node\ \%d,}\)\.{\ req-id\ \%d"},\39\\{size},\39\.{GET%
\_ROOT}(\,)\MG\\{ProcRoot}\MG\\{MyProcID},\39\\{mbox});{}$\6
${}\\{retval}\K\\{PutMessage}(\.{GET\_ROOT}(\,)\MG\\{ProcRoot}\MG\\{MyProcID},%
\39\\{mbox},\39\.{MSG\_TYPE\_USER\_START},\39\T{1},\39{-}\T{1},\39\\{message},%
\39\\{size});{}$\6
\&{if} ${}(\\{retval}<\T{0}){}$\5
${}\{{}$\1\6
\&{char} \\{st}[\T{80}];\7
${}\\{LogError}(\.{EC\_ERROR},\39\.{"RemoteSendMessage"},\39\.{"PutMessage\
error:"});{}$\6
${}\\{sprintf}(\\{st},\39\.{"PutMessage\ on\ node\ }\)\.{\%d:\ "},\39%
\\{GetRoot}(\,));{}$\6
\\{perror}(\\{st});\6
\4${}\}{}$\2\6
\8\#\&{else}\6
${}\\{retval}\K\\{SendNode}(\.{GET\_ROOT}(\,)\MG\\{ProcRoot}\MG\\{MyProcID},\39%
\\{mbox},\39\\{message},\39\\{size});{}$\6
\&{if} ${}(\\{retval}<\T{0}){}$\5
${}\{{}$\1\6
${}\\{printe}(\.{"\%d:\ RemoteSendMessa}\)\.{ge:\\n"},\39\\{GetRoot}(\,));{}$\6
\\{perror}(\.{"Error\ sending\ with\ }\)\.{SendNode"});\6
\4${}\}{}$\2\6
\&{else} \&{if} ${}(\\{retval}>\T{0}){}$\5
${}\{{}$\1\6
${}\\{printe}(\.{"\%d:\ RemoteSendMessa}\)\.{ge:\\n"},\39\\{GetRoot}(\,));{}$\6
${}\\{printe}(\.{"\%d:\ send\ message\ si}\)\.{ze\ \ is\ \ greater\ \ tha}\)%
\.{n\ \ the\ receive\ buffe}\)\.{r\\n"},\39\\{GetRoot}(\,));{}$\6
\4${}\}{}$\2\6
\8\#\&{endif}\6
\4${}\}{}$\2\7
\\{TEXReceiveMessage}(\&{int} \\{mbox}${},\39{}$\&{char} ${}{*}\\{message},%
\39{}$\&{int} ${}{*}\\{size},\39{}$\&{int} \\{timeout})\1\1\2\2\6
${}\{{}$\1\6
\&{int} \\{retval};\6
\8\#\&{ifdef} \.{EPX\_MAILBOXES}\7
${}\\{RR\_Message\_t}*\|m;{}$\6
${}\|m\K\\{malloc}{}$(\&{sizeof} (\\{RR\_Message\_t}));\6
\8\#\&{endif}\6
\&{if} (\\{flag})\1\5
${}\\{LogError}(\.{EC\_ERROR},\39\.{"RecvMessage"},\39\.{"TEXReceiveMessage\ a}%
\)\.{bout\ to\ receive\ a\ ms}\)\.{g\ (size=\%d)\ from\ any}\)\.{\ node\ on\
req-id\ \%d"},\39{*}\\{size},\39\\{mbox});{}$\2\6
\8\#\&{ifdef} \.{EPX\_MAILBOXES}\6
${}\\{retval}\K\\{GetMessage}({-}\T{1},\39\\{mbox},\39\.{MSG\_TYPE\_USER%
\_START},\39{-}\T{1},\39\|m);{}$\6
${}{*}\\{size}\K\|m\MG\\{Header}.\\{Size};{}$\6
${}\\{memcpy}(\\{message},\39\|m\MG\\{Body},\39{*}\\{size});{}$\6
\&{if} ${}(\\{retval}<\T{0}){}$\5
${}\{{}$\1\6
\&{char} \\{st}[\T{80}];\7
${}\\{LogError}(\.{EC\_ERROR},\39\.{"RemoteSendMessage"},\39\.{"GetMessage\
error:"});{}$\6
${}\\{sprintf}(\\{st},\39\.{"GetMessage\ on\ node\ }\)\.{\%d:\ "},\39%
\\{GetRoot}(\,));{}$\6
\\{perror}(\\{st});\6
\4${}\}{}$\2\6
\8\#\&{else}\6
${}{*}\\{size}\K\\{RecvNode}({-}\T{1},\39\\{mbox},\39\\{message},\39{*}%
\\{size});{}$\6
\&{if} (\\{flag})\1\5
${}\\{LogError}(\.{EC\_ERROR},\39\.{"RecvMessage"},\39\.{"TEXReceiveMessage\ r}%
\)\.{eceived\ a\ msg\ on\ req}\)\.{-id\ \%d"},\39\\{mbox});{}$\2\6
\&{if} ${}({*}\\{size}<\T{0}){}$\1\5
${}\\{printe}(\.{"\%d:\ TEXReceiveMessa}\)\.{ge\ error:\ received\ m}\)%
\.{essage\ is\ greater\ th}\)\.{an\ receive\ buffer\\n"},\39\\{GetRoot}(%
\,));{}$\2\6
\8\#\&{endif}\6
\4${}\}{}$\2\7
\\{GetRoot}(\,)\1\1\2\2\6
${}\{{}$\1\6
\&{return} \.{GET\_ROOT}(\,)${}\MG\\{ProcRoot}\MG\\{MyProcID};{}$\6
\4${}\}{}$\2\7
\\{Export}(\&{Alias\_t} \|a${},\39{}$\&{IDF} \|b)\1\1\2\2\6
${}\{\,\}{}$\7
\&{int} \\{TEXGetNumTasks}(\,)\1\1\2\2\6
${}\{{}$\1\6
\&{extern} \&{int} \\{num\_tasks}[\,];\7
\&{return} \\{num\_tasks}[\\{GetRoot}(\,)];\6
\4${}\}{}$\C{ this function should return a value meaning that the task    is
running (\PB{\.{DIR\_RUNNING}}) or waiting for being enabled    (\PB{\.{DIR%
\_WAITING}}). For the time being, as we don't have access    to this
information, we only return \PB{\.{DIR\_RUNNING}}  }\2\7
\.{STATUS}\\{TEXGetTaskStatus}(\&{IDF} \|t)\1\1\2\2\6
${}\{{}$\1\6
\&{return} \.{DIR\_RUNNING};\6
\4${}\}{}$\2\7
\\{TEXStopTask}(\,)\1\1\2\2\6
${}\{{}$\1\6
\\{exit}(\T{1});\6
\4${}\}{}$\2\7
\\{TEXRestartTask}(\,)\1\1\2\2\6
${}\{{}$\1\6
\\{DIRNetGenericComponent}(\,);\6
\\{TEXStopTask}(\,);\6
\4${}\}{}$\2\6
\8\#\&{endif}\par
\U4.\fi

\M{60}This function converts a message-id (as defined in
{\tt dirdefs.h}) into a human-intelligible message.
\Y\B\4\X60:DIR Print Message\X${}\E{}$\6
\&{char} ${}{*}{}$\\{DIRPrintMessage}(\&{int} \|i)\1\1\2\2\6
${}\{{}$\1\6
\&{if} ${}(\|i\G\.{FIRST\_MESSAGE}\W\|i\Z\.{LAST\_MESSAGE}){}$\1\5
\&{return} \\{DIRMessage}${}[\|i-\.{FIRST\_MESSAGE}];{}$\2\6
\&{return} \.{"<unknown>"};\6
\4${}\}{}$\2\7
\&{char} ${}{*}{}$\\{DIRPrintTimeout}(\&{int} \|i)\1\1\2\2\6
${}\{{}$\1\6
\&{switch} (\|i)\5
${}\{{}$\1\6
\4\&{case} \.{IA\_FLAG\_TIMEOUT}:\5
\&{return} \.{"IA\ flag\ timeout"};\6
\4\&{case} \.{MIA\_TIMEOUT}:\5
\&{return} \.{"MIA\ timeout"};\6
\4\&{case} \.{TAIA\_TIMEOUT}:\5
\&{return} \.{"TAIA\ timeout"};\6
\4\&{case} \.{TEIF\_TIMEOUT}:\5
\&{return} \.{"TEIF\ timeout"};\6
\4\&{case} \.{IA\_FLAG\_TIMEOUT\_B}:\5
\&{return} \.{"IA\ flag\ `B'\ timeout}\)\.{"};\6
\4\&{case} \.{MIA\_TIMEOUT\_B}:\5
\&{return} \.{"MIA\ `B'\ timeout"};\6
\4\&{case} \.{TAIA\_TIMEOUT\_B}:\5
\&{return} \.{"TAIA\ `B'\ timeout"};\6
\4\&{case} \.{TEIF\_TIMEOUT\_B}:\5
\&{return} \.{"TEIF\ `B'\ timeout"};\6
\4\&{case} \.{IAT\_TIMEOUT}:\5
\&{return} \.{"IA\ Task\ timeout"};\6
\4\&{case} \.{INJECT\_FAULT\_TIMEOUT}:\5
\&{return} \.{"F.\ Injecting\ timeou}\)\.{t"};\6
\4\&{default}:\5
\&{return} \.{"<unknown>"};\6
\4${}\}{}$\2\6
\4${}\}{}$\2\7
\&{char} ${}{*}{}$\\{DIRPrintCode}(\&{int} \|i)\1\1\2\2\6
${}\{{}$\1\6
\&{char} ${}{*}\|s;{}$\7
${}\|s\K\\{DIRPrintTimeout}(\|i);{}$\6
\&{if} ${}(\\{strcmp}(\|s,\39\.{"<unknown>"})\E\T{0}){}$\1\5
\&{return} \\{DIRPrintMessage}(\|i);\2\6
\&{return} \|s;\6
\4${}\}{}$\2\7
\&{char} ${}{*}{}$\\{role2ascii}(\&{int} \\{role})\1\1\2\2\6
${}\{{}$\1\6
\&{switch} (\\{role})\5
${}\{{}$\1\6
\4\&{case} \.{DIR\_MANAGER}:\5
\&{return} \.{"manager"};\6
\4\&{case} \.{DIR\_BACKUP}:\5
\&{return} \.{"backup\ agent"};\6
\4\&{default}:\5
\&{return} \.{"<unknown>"};\6
\4${}\}{}$\2\6
\4${}\}{}$\2\par
\U4.\fi

\M{61}The I'm Alive Task.
After initialisation time, the IAT enters a waiting status in which
it waits for an activation message from the agent it guards...
\Y\B\4\X61:I'm Alive Task\X${}\E{}$\6
\&{message\_t} \\{amessage};\6
\&{timeout\_t} \\{aia};\6
\&{TOM} ${}{*}\\{atom};$ \X63:The \PB{\\{IAlarm}} function\X\6
\X62:The \PB{\\{DIRAlive}} function\X\par
\U4.\fi

\M{62}This is the code for the I'm Alive Task.
\Y\B\4\X62:The \PB{\\{DIRAlive}} function\X${}\E{}$\6
\&{int} \\{DIRAlive}(\&{void})\1\1\2\2\6
${}\{{}$\1\6
\&{int} \\{subid};\6
\&{int} \\{IAlarm}(\&{TOM} ${}{*});{}$\7
${}\\{atom}\K\\{tom\_init}(\\{IAlarm});{}$\6
\4\\{restart}:\5
${}\\{LogError}(\.{EC\_ERROR},\39\.{"IAT"},\39\.{"IAT\ starts..."});{}$\6
\X64:wait for activation message\X\6
${}\\{LogError}(\.{EC\_ERROR},\39\.{"IAT"},\39\.{"IAT\ activated\ ..."});{}$\6
${}\\{subid}\K\\{amessage}.\\{subid}{}$;\C{ the alarm of these timeouts simply
sends a message of id 	   <timeout-type>, subid = <subid> to the IAT 	 }\6
\X65:insert timeout (IA-flag-timeout, CYCLIC, \PB{\\{subid}})\X\6
${}\\{LogError}(\.{EC\_ERROR},\39\.{"IAT"},\39\.{"timeout\ activated,\ }\)%
\.{now\ enter\ the\ main\ l}\)\.{oop..."});{}$\6
\&{while} (\T{1})\5
${}\{{}$\1\6
${}\\{LogError}(\.{EC\_ERROR},\39\.{"IAT"},\39\.{"waiting\ for\ a\ messa}\)%
\.{ge"});{}$\6
\X66:IATask: wait for an incoming message\X\6
${}\\{LogError}(\.{EC\_ERROR},\39\.{"IAT"},\39\.{"a\ message\ came\ in\ :}\)\.{%
\ \%d\ (\%s)"},\39\\{amessage}.\\{type},\39\\{DIRPrintCode}(\\{amessage}.%
\\{type}));{}$\6
\&{if} ${}(\\{amessage}.\\{type}\E\.{IAT\_TIMEOUT}){}$\5
${}\{{}$\C{ time to check the IA-flag! }\1\6
\&{if} ${}(\\{IA\_flag}\E\T{0}){}$\5
${}\{{}$\1\6
${}\\{LogError}(\.{EC\_ERROR},\39\.{"IAT"},\39\.{"IA\ flag\ is\ zero,\ co}\)%
\.{rrect."});{}$\6
${}\\{IA\_flag}\K\T{1};{}$\6
${}\\{LogError}(\.{EC\_ERROR},\39\.{"IAT"},\39\.{"IA\ flag\ has\ been\ se}\)%
\.{t\ again."});{}$\6
\4${}\}{}$\2\6
\&{else}\5
${}\{{}$\1\6
${}\\{LogError}(\.{EC\_ERROR},\39\.{"IAT"},\39\.{"A\ L\ A\ R\ M!!!"});{}$\6
\X67:send \PB{\.{TEIF}} to all except \PB{\\{subid}}\X\6
\X68:delete timeout (IA-flag-timeout)\X\6
\&{break};\6
\4${}\}{}$\2\6
\4${}\}{}$\2\6
\4${}\}{}$\C{ end loop }\C{ TEXRestartTask(); }\2\6
\&{goto} \\{restart};\6
\4${}\}{}$\C{ end alive }\2\par
\U61.\fi

\M{63}The Alarm function for the I'm Alive Task.
\Y\B\4\X63:The \PB{\\{IAlarm}} function\X${}\E{}$\6
\&{int} \\{IAlarm}(\&{TOM} ${}{*}\\{atom}){}$\1\1\2\2\6
${}\{{}$\1\6
${}\\{amessage}.\\{type}\K\.{IAT\_TIMEOUT};{}$\6
${}\\{TEXSendMessage}(\.{IAT\_MBOX},\39{}$(\&{char} ${}{*}){}$ ${}{\AND}%
\\{amessage},\39{}$\&{sizeof} (\\{amessage}));\6
\4${}\}{}$\2\par
\U61.\fi

\M{64}The I'm Alive Task here is waiting for an activation message
from its generic component.
\Y\B\4\X64:wait for activation message\X${}\E{}$\6
${}\{{}$\1\6
\&{int} \\{size}${}\K\&{sizeof}(\&{message\_t});{}$\7
${}\\{LogError}(\.{EC\_ERROR},\39\.{"<wait\ for\ activatio}\)\.{n\ message>"},%
\39\.{"waiting\ for\ message}\)\.{s..."}){}$;\C{flag=1;}\6
${}\\{TEXReceiveMessage}(\.{IAT\_MBOX},\39{}$(\&{char} ${}{*}){}$ ${}{\AND}%
\\{amessage},\39{\AND}\\{size},\39\.{INFINITE});{}$\6
${}\\{LogError}(\.{EC\_ERROR},\39\.{"<wait\ for\ activatio}\)\.{n\ message>"},%
\39\.{"got\ a\ message\ (\%s)"},\39\\{DIRPrintCode}(\\{amessage}.\\{type}));{}$%
\6
\4${}\}{}$\2\par
\U62.\fi

\M{65}Also the I'm alive thread makes use of the timeout manager.
\Y\B\4\X65:insert timeout (IA-flag-timeout, CYCLIC, \PB{\\{subid}})\X${}\E{}$\6
${}\{{}$\1\6
\&{int} \\{IAlarm}(\&{TOM} ${}{*});{}$\7
${}\\{tom\_declare}({\AND}\\{aia},\39\.{IAT\_CYCLIC},\39\.{TOM\_SET\_ENABLE},%
\39\.{IAT\_TIMEOUT},\39\\{subid},\39\.{IAT\_DEADLINE});{}$\6
${}\\{tom\_insert}(\\{atom},\39{\AND}\\{aia});{}$\6
\4${}\}{}$\2\par
\U62.\fi

\M{66}Same as $<$wait for an incoming message$>$, but for the I'm Alive Task.
\Y\B\4\X66:IATask: wait for an incoming message\X${}\E{}$\6
${}\{{}$\1\6
\&{int} \\{size}${}\K\&{sizeof}(\&{message\_t});{}$\7
${}\\{TEXReceiveMessage}(\.{IAT\_MBOX},\39{}$(\&{char} ${}{*}){}$ ${}{\AND}%
\\{amessage},\39{\AND}\\{size},\39\.{INFINITE});{}$\6
${}\\{LogError}(\.{EC\_ERROR},\39\.{"IAT"},\39\.{"got\ message\ \%d\ (\%s)}\)%
\.{"},\39\\{amessage}.\\{type},\39\\{DIRPrintCode}(\\{amessage}.\\{type}));{}$\6
\4${}\}{}$\2\par
\U62.\fi

\M{67}
\Y\B\4\X67:send \PB{\.{TEIF}} to all except \PB{\\{subid}}\X${}\E{}$\6
${}\{{}$\1\6
\&{int} \|i;\6
\&{int} \\{you}${}\K\\{GetRoot}(\,);{}$\7
\&{for} ${}(\|i\K\T{0};{}$ ${}\|i<\.{MAX\_PROCS};{}$ ${}\|i\PP){}$\5
${}\{{}$\1\6
\&{if} ${}(\|i\I\\{subid}\W\|i\I\\{you}){}$\5
${}\{{}$\1\6
${}\\{amessage}.\\{type}\K\.{TEIF}{}$;\C{ i.e., ``ENable IAt'' }\6
${}\\{amessage}.\\{subid}\K\\{GetRoot}(\,);{}$\6
${}\\{RemoteSendMessage}(\|i,\39\.{ALIAS}(\|i),\39{}$(\&{char} ${}{*}){}$ ${}{%
\AND}\\{amessage},\39{}$\&{sizeof} (\\{amessage}));\6
\4${}\}{}$\2\6
\4${}\}{}$\2\6
\4${}\}{}$\2\par
\U62.\fi

\M{68}
\Y\B\4\X68:delete timeout (IA-flag-timeout)\X${}\E{}$\6
$\\{tom\_delete}(\\{atom},\39{\AND}\\{aia}){}$;\par
\U62.\fi

\M{69}
\Y\B\4\X69:Spawn the I'm Alive Task\X${}\E{}$\6
${}\{{}$\1\6
\&{int} \\{DIRAlive}(\&{void});\7
${}\\{StartThread}(\\{DIRAlive},\39\T{65536},\39{\AND}\\{errors},\39\T{0},\39%
\NULL);{}$\6
\4${}\}{}$\2\par
\U6.\fi

/* eof dirnet.w */

\inx
\fin

\vskip1cm

\noindent
{\bf References}

[1] V. De Florio, G. Deconinck, R. Lauwereins. An Algorithm for Tolerating
Crash Failures in Distributed Systems.
In Proc. of the 7th Annual IEEE International
Conference and Workshop on the Engineering of Computer Based
Systems (ECBS), Edinburgh, Scotland, 3--5 April 2000. IEEE.

[2] V. De Florio, G. Deconinck, R. Lauwereins. The {EFTOS} Voting Farm: a
Software Tool for Fault Masking in Message Passing Parallel Environments.
In Proc. of the 24th Euromicro Conference (Euromicro '98),
Workshop on Dependable Computing Systems, V{\"a}ster\aa{}s, Sweden, August
1998. IEEE.

[3] V. De Florio, G. Deconinck, R. Lauwereins. Software Tool Combining Fault
Masking with
User-defined Recovery Strategies. {\it IEE Proceedings -- Software\/} {\bf 145}(6),
1998. IEE.

[4] V. De Florio, C. Blondia.
Dynamics of a Time-outs Management System.
{\it Complex Systems\/} {\bf 16}(3), 2006.
Complex Systems Publications, Champaign, IL.

[5] V. De Florio, C. Blondia.
Design Tool To Express Failure Detection Protocols. {\it IET Software\/} {\bf 4}(2),
2010. IET.

\con